\documentclass[notitlepage,12pt]{jedm}
\usepackage[table]{xcolor}
\usepackage{graphicx}
\usepackage{subfigure}
\usepackage{url}
\usepackage{hyperref}
\usepackage{multirow}
\usepackage{hhline}
\usepackage{amsmath}
\usepackage{mathpazo} 
\usepackage{todonotes}

\hypersetup{
  colorlinks   = true, 
  urlcolor     = blue, 
  linkcolor    = blue, 
  citecolor   = blue 
}

\begin{document}

\title{Student-centric Model of Learning Management System Activity and Academic Performance: from Correlation to Causation}
\date{} 

\author{{\large Varun Mandalapu }\\University of Maryland Baltimore County\\varunm1@umbc.edu \and {\large Lujie Karen Chen}\\University of Maryland Baltimore County\\lujiec@umbc.edu  \and {\large Sushruta Shetty}\\University of Maryland Baltimore County\\sshetty2@umbc.edu  \and {\large Zhiyuan Chen}\\University of Maryland Baltimore County\\zhchen@umbc.edu \and {\large Jiaqi Gong}\\Univerisy of Alabama\\jiaqi.gong@ua.edu}

\maketitle

\begin{abstract}

In recent years, there is a lot of interest in modeling students' digital traces in Learning Management System (LMS) to understand students' learning behavior patterns including aspects of meta-cognition and self-regulation, with the ultimate goal to turn those insights into actionable information to support students to improve their learning outcomes. In achieving this goal, however, there are two main issues that need to be addressed given the existing literature. Firstly, most of the current work is course-centered (i.e. models are built from data for a specific course) rather than student-centered (i.e. models are built taking the perspective of students by analyzing data across courses); secondly, a vast majority of the models are correlational rather than causal. Those issues make it challenging to identify the most promising actionable factors for intervention at the student level where most of the campus-wide academic support is designed for. In this paper, we explored a student-centric analytical framework for LMS activity data that can provide not only correlational but causal insights mined from observational data. We demonstrated this approach using a dataset of 1651 computing major students at a public university in the US during one semester in the Fall of 2019. This dataset includes students' fine-grained LMS interaction logs and administrative data, e.g. demographics and academic performance. In addition, we expand the repository of LMS behavior indicators to include those that can characterize the time-of-the-day of login (e.g. chronotype). Our analysis showed that student login volume, compared with other login behavior indicators, is both strongly correlated and causally linked to student academic performance, especially among students with low academic performance. We envision that those insights will provide convincing evidence for college student support groups to launch student-centered and targeted interventions that are effective and scalable. 

{\parindent0pt
\textbf{Keywords:} Learning Management System, Student behavior modelling, Causal Analysis}
\end{abstract}

\section{Introduction}


Understanding student learning behaviors is a complex process as it is affected by a multitude of factors like self-regulation, instructional design, and social environment \cite{ikink2019vu}. However, it is not efficient to study all the influencing factors as some factors are not feasible to intervene compared to others. Some have a higher impact on the performance at an individual level than others. Earlier research in education identified self-regulation as a highly impactful and intervenable factor \cite{zimmerman2014comparing,schapiro2000dynamic}. Self-regulation, in general terms, is defined as one's ability to control their own behaviors, thoughts, and emotions to achieve their goals to drive their learning experiences successfully.

\setlength{\parindent}{0em}
\setlength{\parskip}{1ex}

Studying student self-regulation capabilities is an arduous task as it involves multidimensional and complex factors like planning, goal setting, time management, and self-monitoring. Many of these factors are hard to quantify without student self-reports or other psychometric evaluations, and these reports are prone to biases like social desirability and reference bias \cite{toering2012measuring,rosenman2011measuring}. To mitigate these biases, researchers in education focused on extracting data from computer-based systems like Learning Management Systems (LMS) as they are used to deliver course content and have the capability to capture student interaction and assessment behaviors non-invasively \cite{tempelaar2015search,salehian2021measuring}. Student interaction data with LMS proved to be a valuable resource in identifying learning strategies and track patterns among students that strongly support their academic performance. The data logs collected by LMS systems are analyzed with scientific techniques published in the Educational Data Mining (EDM) domain. In their study, Romero and Ventura \cite{romero2007educational} described that EDM methods rely on clustering and pattern recognition techniques to categorize students into various groups based on their interaction patterns. Categorization of students using clustering and pattern recognition supports instructors in making changes for a set of students. Teaching practices that impact the entire classroom can be evaluated using predictive analytics that tracks student learning and achievement from the vast amount of interaction data collected by LMS. Majority of these studies analyze student interaction data in a specific course to understand student learning behaviors and teaching methods.

While course level predictions are suitable for supporting instructor level decision making; however, if intervention is on student level behaviors such as study habits or self-regulation skills, it is beneficial to look at student-centered indicators so that interventions may be more targeted and cost-effective. Developing student-centric models that analyze student LMS interactions across courses in a college/university setting will help address the issues with course specific models. This study is the first step in developing feature extraction and modeling methods from LMS login data that are scalable across semesters and transferable across different undergraduate fields. Developing these early performance prediction models alone does not help improve student learning as identifying self-regulation components like time management and other student engagement factors play a crucial role in recommending methods to improve their current behaviors.

LMS systems also enabled researchers to study time management strategies, a component of self-regulation adopted by students \cite{jo2015constructing,ahmad2020analytics}. In contradiction to traditional time management studies that utilize self-reported questionnaires, student time management strategies are analyzed based on login and submission patterns in LMS systems. However, only few studies focused on the relationship between time management strategies identified in LMS systems with the biological nature of human functionality. Therefore, these earlier modeling efforts are limited to predicting student performance as a function of their login patterns and fell short of identifying patterns in time management that support improving student academic improvement. These earlier studies also focused on student behaviors related to their time compliance with assignments, including procrastination behaviors and ways to support the earliness of student work \cite{ilves2018supporting,edwards2009comparing,martin2015effects}. These studies reported that early starters have better learning outcomes, but there is little emphasis on what times these students work during a day and how these learning time patterns affect their performance. Inspired by these earlier studies and comparatively similar studies in biology about human activity/productivity during different times of the day, referred to as chronotype/chronological analysis, this study analyzes the relationship between student activity on LMS at different times in a day and their performance by employing clustering methods. In addition to this, the student chronotype clusters are added to LMS behaviors and demographics as features to predictive models to predict student performance.

Earlier research in the area of Learning Analytics and Educational Data Mining focused on the relationship between self-regulation and student performance through LMS systems used student interaction features as proxy variables for self-regulation \cite{dabbagh2013using,landrum2020examining}. These studies showed a significant correlation between student login patterns and their academic performance. However, it is still challenging to design interventions as correlations do not necessarily mean a meaningful cause and effect between student login behaviors and academic performance. In this work, we introduce a novel framework to perform causal inference and discovery of multimodal student data collected from LMS and student administrative system. In addition, we enrich the feature set by introducing new approach to extract login behaviors at a student level. Finally, this work also introduced concepts from human chronobiology to analyze student login behavior on an hourly basis (Chronotype analysis) to identify activity patterns at different times in a day and their relationship to academic performance.  The proposed correlation/predictive modeling framework \cite{mandalapu2021student}, chronotype analysis and causal inference framework goes beyond traditional educational data mining approaches to answer the following research questions that are under-studied in current literature.

\section{Research Questions \& Outline}

RQ1. How student learning activity at different times in a day (chronotypes) relate to their academic performance?

RQ2. How student-centric characterization of LMS activity correlates with academic performance? 

RQ3. What are the causal relationship between student-centric characterization of LMS activity and academic performance? 

In the related work Section \ref{Related Work}, we discuss earlier research works that use LMS system data to predict student academic achievement. Section \ref{Related: CircadianRhythmsChronotypes}, we details research work that focuses on the evolution of chronotype on the basis of human biology and in the following Section \ref{Related: Chronotypes and their relationship with Cognitive abilities and academic achievement} explains how chronotypes are linked to individual cognitive abilities and academic achievement.The final related work sections \ref{Related: Correlation study of the relationship between LMS activity and academic performance} and \ref{Related: Causal inference from Student LMS Interaction Data in Performance Predictions} explains earlier research work in correlation and causation to identify relationship between student LMS interaction data and their academic performance.

Section \ref{DataSet} provides the details of data set used to perform analysis in this study. The methodology section \ref{Methodology} details the steps taken to extract student aggregate login features in \ref{Method:feature extraction}. The correlation studies that focuses on machine learning modeling, and post hoc LIME based explanations are detailed in Section \ref{Method: Machine Learning approach} and Section \ref{Method: Model Explanation using LIME}. The section \ref{Method: causal-inference} details the causal analysis and inference tools and methods adopted in this study. The results of the study related to student chronotype analysis are listed in section \ref{Result: Chronotypes Analysis}. Section \ref{Result: Predictive Model Performance} explains the results of predictive models by comparing their predictions based on multiple evaluation metrics, this section also details various correlation based explanation methods that details the relationship between student demographics, LMS interaction features and academic performance. The causal discovery and inference results are detailed in section \ref{Result: Causal Analysis}. Finally in Section \ref{Discussion}, we detail the key findings and contributions made by this study to the existing literature in subsections \ref{Disc: Key Findings} and \ref{Disc: Key Contributions}.

\section{Related Work and Background}
\label{Related Work}

Universities and colleges around the world adopted LMS systems, such as Moodle and Blackboard, to provide onsite, hybrid, and online courses based on their capabilities to support communication, content creation, administration, and assessment\cite{alokluk2018effectiveness,berechetroad} LMS systems build and swiftly distribute individualized learning materials and information in addition to automating and centralizing a variety of administrative processes including setting up and managing student accounts, establishing a syllabus, assignments, assessments, and grading, etc.\cite{shchedrina2021providing}. These methods also encourage the reuse of instructional materials. Instructors can design content structures, distribute it in a sequential order, limit access, organize group activities, monitor student activity, load and replace learning materials, and give feedback on assessments using the systems. LMS systems use a variety of login roles based on user classification thanks to cutting-edge database software created by Oracle, IBM, and Microsoft that emphasizes interconnection, data independence, and security. These roles will permit instructors to create new content or privately address student issues and create discussion boards to capture student knowledge on specific topics.


The relationship between the use of LMS and student academic accomplishments has been examined in numerous papers in the fields of Educational Data Mining and Learning Analytics. According to Vengroff and Bourbeau's \cite{vengroff2006class} study, undergraduate students benefited from having more content available in LMS. They also conclude that students who used LMS regularly did better in exams than their peers who have minimal interactions. In their study, Dutt and Ismail \cite{dutt2019can} found that keeping track of the LMS resources that students use encourages the creation of fresh approaches to learning that improve student achievement. Additionally, they examined the thresholds for student involvement elements like self-assessment quizzes, exercise time, discussion boards, and performance results. Another study by Lust et al. \cite{lust2011tool} investigated how differently students used various LMS capabilities, such as the amount of time spent on web-links, web lectures, quizzes, feedback, discussion board entries, and messages read. The findings of this research made a significant contribution to the creation of adaptive and creative recommendation systems. In their research, Hung and Zhang \cite{hung2012examining} also discovered trends based on six indices that represented student effort: the frequency with which students access course materials, the number of LMS logins, the sum of interactions in discussion threads, the number of synchronous discussions, the number of posts read, and the course's final grades.

When investigating the relationship between students' online behavior on the LMS and their grades, Dawson et al. \cite{dawson2010harnessing} found a substantial difference between high and low performing students in the number of online sessions visited, overall time spent, and the number of posts in discussion boards. A multinomial logistic regression model was created by Damainov et al. \cite{damianov2009time} based on the amount of time spent in the LMS. This study discovered a substantial correlation between the amount of time students spent studying and their grades, particularly for those who had grades between D and B. Other studies emphasized how frequently students accessed course materials in the LMS, as opposed to measuring time spent online. A study by Baugher et al. \cite{baugher2003student} found that regularity in student hits is a reliable predictor of student performance compared to the total number of hits. In their study, Chancery and Haque studied the student interaction logs of 112 undergraduate students and discovered that those with lower LMS access rates had lower grades than their counterparts with greater access rates. Biktimirovan and Klassen \cite{biktimirov2008relationship}, who observed a significant correlation between student hit consistency and success, provided additional support for this study. In their study, which measured access to numerous LMS features, it was discovered that the only significant predictor of student achievement is access to assignment solutions. However, these studies tend to be more descriptive than predictive.

\subsection{Circadian Rhythms \& Chronotypes}
\label{Related: CircadianRhythmsChronotypes}
Humans have a 24-hour internal clock running in the background that determines when to sleep and when to be productive. These 24-hour cycles are referred to as circadian rhythms. The behavioral manifestation of these circadian rhythms is termed chronotype \cite{reppert2002coordination,dibner2010mammalian}. Understanding an individual’s chronotype helps identify their routine and provides insights about their highly active and productive time. Earlier research argues that circadian rhythms differ between individuals as a group of clock genes conditions them \cite{bell2005circadian}. Nonetheless, these are not fixed and can vary during an individual’s lifetime.

Students have to schedule their daily activities based on their predetermined schedules by taking social constraints like class schedules, outside work hours into consideration. General clock time preferences are more likely to match natural chronotypes when students are given more flexibility in organizing their schedules. Differences between a person’s natural chronotypes and schedules influenced by social constraints cause a phenomenon referred to as social jetlag \cite{roenneberg2003life,wittmann2006social}. This social jetlag leads to an accumulation of sleep debt that causes tiredness and a decline in cognitive abilities throughout the workweek. Prior research suggested that individuals with a match in chronotypes and schedules during workweek don’t show a change on weekends \cite{korczak2008influence,vitale2015chronotype}. This is in contrast with individuals whose weekday routine varies from natural routines. As activity and sleep are both indicators of natural chronotypes, an observation of individuals on weekends might give accurate insights into their natural habits defined by circadian rhythms.

Earlier studies primarily focused on two significant chronotypes in humans: Morning and Evening. These studies showed that these two chronotypes are based on an individual’s age and gender and vary across an individual lifetime. The insights from earlier work showed that children are predisposed towards morningness, but a delay of phase preference can be observed when they reach adolescence \cite{carskadon1998adolescent,crowley2007sleep}. This shift reaches a maximum of eveningness at the age of 20. Multiple studies also observed that most individuals return to morningness again at the age of 50 \cite{diaz2008morningness,baehr2000individual}. Gender-based chronotype studies contradicted a lot in identifying variations in morning and eveningness as some studies showed women to have a greater tendency towards morningness compared to men. A meta-analysis showed that the significance between morningness and females is very weak \cite{randler2007gender}. In addition to these two demographics, other researches focus on different variables like productivity, mood, temperament, caffeine consumption, avocational interest, and internal temperatures.

Recent studies observed a new type of chronotype referred to as intermediate \cite{valladares2018individual,porcheret2018chronotype}. A study performed by Putilov et al. in 2019 \cite{putilov2019there} argued about four chronotypes based on time differences in alertness and sleepiness: morning, afternoon, napper, and evening. Most of these studies adopt different self-reported questionnaires that are developed to analyze an individual’s daily preferences. Initially, most of these measures treated chronotypes as unidimensional: Diurnal Type Scale (DTS) and Circadian Composite scale (CCS) \cite{torsvall1980diurnal,smith1989evaluation}. But, psychometric studies questioned this unidimensional approach to morning-eveningness \cite{putilov2005sleepless,putilov200552}. Recent studies came up with a multidimensional approach that treated chronotypes as independent dimensions. As most of the chronotype measurement methods are self-reported surveys, biases like social desirability, recall period, sampling approach, and selective recall are hard to eliminate. In order to mitigate this issue, our work explores the activity patterns of students based on login times captured by the LMS. We employ clustering techniques to identify and classify students based on their similar activity patterns.

\subsection{Chronotypes and their relationship with Cognitive abilities and academic achievement}
\label{Related: Chronotypes and their relationship with Cognitive abilities and academic achievement}
Research in psychology documented the effects of time of day on complex simple and complex cognitive functions that are based on an individual’s chronotype \cite{schmidt2007time,roberts1999morningness}. An early study by Roberts and Kyllonen \cite{kyllonen1990reasoning} showed that individuals who were active during the evenings did well on the measures of memory, cognitive ability, and processing speeds even though these cognitive tasks are performed during the morning time. In addition to this, this study also reported a high correlation between individuals working memory which is a proxy of general intelligence, and morningness and eveningness \cite{killgore2007morningness}. However, other researchers reported that the relationship between cognitive ability and chronotype is much more variegated. For example, a study by Killgore and Killgore\cite{killgore2007morningness} reported a significant correlation between eveningness and verbal cognitive ability but not between eveningness and math ability. Further analysis showed that the latter finding is only observed in female participants. A similar study by Song and Stough\cite{song2000relationship} showed a significant eveningness effect on a spatial subtest of Multidimensional Aptitude Battery IQ and not on any other subtests. Most of the research in this domain is still inconsistent on which aspects of cognitive abilities do chronotypes impact.

Inspired by earlier studies that reported significant relations between cognitive ability and chronotypes, researchers in education focused on studying the relationship between academic performance and chronotypes, mainly student grade point averages and other measures extracted from their academic achievement indicators \cite{enright2017chronotype,preckel2011chronotype}. Most of the studies in this area reported that student academic achievement is strongly and inversely related to their eveningness, whereas morningness is positively associated with their achievement. These patterns are similar in students at the high school level and students at the university level. A study by Precckel and Roberts \cite{preckel2009schulleistung} showed a significant negative correlation between academic achievement and eveningness. This study was conducted on 270 German secondary students where teachers assigned grades averaged over German, Math, English, and Physics. These results seem to be consistent even after introducing controls for gender and intelligence.

A recent study focused on university students that utilized three classification scales (Morning, Evening, and Neither) found that morningness students showed much better performance on both theoretical and practical examples compared to students belonging to other chronotypes \cite{montaruli2019effect}. This study also showed that students belonging to neither evening nor morning chronotypes performed lower than their counterparts. Additionally, this study also observed that students who belong to eveningness did worse on practical examinations than morningness and students belonging to neither class. One interesting find in this study is related to the performance of eveningness students on theoretical exams. This study infers that the higher intelligence expressed by eveningness students compensated for their disadvantage on the theoretical exam but not on practical exams. However, it is still necessary to study the impact of an age-based shift in chronotypes with student’s academic performance, as earlier studies reported that individuals move from morningness to eveningness by the time they reach their 20’s. In addition to this, it is also necessary to study the impact of chronotypes on student-centric models that analyze comprehensive student data in a semester or academic year instead of course-specific analysis. This understanding will support the development of interventions that effective and promise overall student success. Our work will utilize student clusters extracted from their hourly login patterns to develop and assess models that predict student performances.

\subsection{Correlation study of the relationship between LMS activity and academic performance} 
\label{Related: Correlation study of the relationship between LMS activity and academic performance}
Online teaching tactics are mostly dependent on instruction design since each type of interaction—student/instructor, student/student, and student/content—has a favorable effect on learning. A study by Coldwell et al. \cite{coldwell2008online} focused on the relationship between student participation in a fully online course and their final grades. They discovered a positive correlation between student participation and final grade. When Dawson et al. \cite{dawson2010harnessing} looked at the effects of several LMS technologies, they discovered a positive correlation between participation in discussion forums and student achievement. They found that the discussion forum, which is the main instrument for interaction in LMS, was where more than 80\% of interactions took place. Asynchronous communication was discovered to be the main method of engagement in all online courses, according to a different study by Greenland \cite{greenland2011using}. Near the deadlines for assignments and exams, Nandi et al. \cite{nandi2011active} discovered an increase in the volume of posts in discussion forums. Additionally, they discovered a strong relationship between exam performance and ongoing online class engagement, particularly among top achievers.

All of the studies mentioned used log files from LMS systems to gather objective information about activity and performance in order to establish a connection between independent interaction variables and student grades. The majority of the studies that have been addressed are based on univariate analysis and concentrate on one variable, or a group of variables that have a significant impact on the outcomes of students in one course or courses like it. To assess or comprehend student performance, especially across the range of on-campus courses available in a university context, is a very difficult task.  Most of the authors discussed above noted the need for more in-depth works to investigate student performance across courses and based on multiple variables. These studies also lack an explanation of the variables they used to measure student performance, and it is clear that the authors chose the LMS variables based on their belief that these variables are highly correlated with student scores.

\setlength{\parindent}{0em}
\setlength{\parskip}{1ex}

Even though there is a common agreement about the purpose of learning analytics, there are still several varying opinions on what data needs to be collected and analyzed to improve teaching and learning processes. It is extremely difficult to determine the net contribution of numerous interactions to the learning processes, according to a study by Agudo-Peregrina et al. \cite{agudo2014can}. Contrary to past studies that indicated a strong importance for student peer connections, their findings indicate that peer contact between students has less influence than peer interaction between students and teachers. In order to forecast student grades at the end of the course, Dominquez et al. \cite{dominguez2016predicting} used a variety of factors, including LMS logins, time stamps, and content access flags recorded in a biology course. According to the findings, the algorithm's forecast accuracy for succeeding semesters is 50\%. In their recent study, Lerche and Keil \cite{lerche2018predicting} used Moodle log data from 369 students enrolled in three online courses over the course of three semesters to forecast their final grades for each course. For all three courses, their regression findings ranged from 0.17 to 0.6 in terms of predicting student scores in a course at the end of the semester. This wide variation in performance between courses is caused by the different variables used in each course according to its structure. The inconsistent results from past studies may be explained by examining the different instructional designs, factors in extracted data, statistical inferences, predictive modeling employed, interpreting model outcomes and pattern observations, etc.

\setlength{\parindent}{0em}
\setlength{\parskip}{1ex}

Since LMS systems record student interactions in non-intrusive, ready-to-use settings, their data has gained prominence in LA and EDM circles. Several studies were discussed earlier in this research that utilized the LMS data to develop models that track student progress. However, it is currently difficult to develop extremely precise models that forecast student learning outcomes across courses and understand the impact of various variables gathered by LMS. Their failure to predict student success across courses in a particular semester is another important flaw in past research. One primary issue in predicting student performance across a semester is to find methods that aggregate student LMS variables across courses. This study demonstrates approaches for filling the knowledge gap identified by past investigations.

\setlength{\parindent}{0em}
\setlength{\parskip}{1ex}

In this study, we approach the problem of tracking student achievement by developing student-centric models that build on aggregated LMS interaction variables collected across a semester irrespective of student year and course. One distinctive feature of our  work is related to the study of model performance on longitudinal student data. We develop models that predict student end-of-term GPA based on four cumulative periods in a semester. The purpose of this paper is to explain how numerous aggregated LMS variables affect various student groups that have been grouped according to performance, race, gender, and student type. The importance of features is explained by adopting correlation statistics for univariate importance, a regression model for interaction effect, and LIME for model-based yet model agnostic explanations.

\subsection{Causal inference from Student LMS Interaction Data in Performance Predictions}
\label{Related: Causal inference from Student LMS Interaction Data in Performance Predictions}

The primary objective of educational research is to develop interventions that promote student academic achievement. Earlier studies in learning analytics and educational data mining focused on developing models that predict student achievements early on in their academic program to support the development of these interventions. However, most of these studies focused on course-level predictions and analysis, making it harder to develop interventions at the student level. Another gap observed in prior research is the lack of causal understanding related to features that causes a shift in student academic performance. Causal inferences promote the development of impactful intervention techniques that positively impact student academic performance.

Traditional research utilizes randomized control trials to discover causal knowledge. However, these are time-consuming, expensive, and sometimes unethical and impractical to implement with real-world students. Studying causal relationships from student interaction with LMS systems allows us to explore a vast amount of noninvasive data collected by these systems. This method will provide a search for answers beyond common correlations. Yet, this type of analysis is not common in learning analytics and educational data mining domains. For example, a study by Fanscali \cite{fancsali2015confounding} investigated the role of carelessness in analyzing the counter intuitive relationship between the learning outcomes of students and their affective states related to confusion and boredom. This study utilized observational data from the algebra cognitive tutor platform. An earlier study by Fanscali \cite{fancsali2014causal} on a similar dataset showed a relationship between student’s affective states, gaming the system, prior knowledge, and their learning outcomes. This study utilized a causal framework termed causal discovery with model. In their research, Koedinger et al. \cite{koedinger2016doer} analyzed interaction data captured by the online learning environment. This study reported causal interaction between active student engagement and their learning outcomes. Inspired by this work, a recent study by Chen et al. \cite{chen2021affect} developed a causal discovery framework that utilized TETRAD \cite{ramsey2018tetrad}, a causal discovery and inference toolkit. Our work adopts the framework proposed by Chen et al. \cite{chen2021affect} to study causal relationships between student login behaviors captured by LMS and learning outcomes.

\section{DataSet}
\label{DataSet}
For this study, we chose undergraduate student data captured by LMS in Fall 2019 from a large public university in the United States. These students were part of either Information Systems (IS) or Computer Science (CS) departments. The Blackboard system is predominantly used as an LMS to deliver course material, assessment, and grading. The student demographic data captured by a standalone Student Information System (SIS) is used to categorize students based on different demographic variables. A total of 1651 students were enrolled in these two departments in the Fall 2019 semester. Based on student distribution in the undergraduate majors selected, we categorized students into three ethnicities based on their counts: White, Asian, and Minority(mostly African Americans and Hispanic population). This study also researches student performance based on their admit types, such as four-year regular student or transfer student. The transfer student here refers to students that completed at least one year of their undergraduate course work at a different college (mostly community college) before enrolling at current institution. The demographics of student data are provided in Table~\ref{tab:1}. The university's Institutional Review Board (IRB) approved this study, and all the student specific demographic and personal information are anonymized by following General Data Protection Regulation (GDPR) standards.

\begin{table}[!ht]
  \caption{Student demographics.}\vspace*{1ex}
  \label{tab:1}
  \centering 
  \begin{tabular}{ c  c } 
  \hline
    \multicolumn{1}{c}{\textbf{Demographic}} & \multicolumn{1}{c}{\textbf{Student Count}} \\
    \hline
    Total Students (N) & 1651\\
    \setlength{\parskip}{1ex}\\
    No of unique courses & 440\\
    \setlength{\parskip}{1ex}\\
    No of unique course instructor combinations & 638\\
    \setlength{\parskip}{1ex}\\
    Male : Female & 1302 (79\%) : 369 (21\%)\\
    \setlength{\parskip}{1ex}\\
    White : Asian : Minority & 630 (38\%) : 495 (30\%) : 526 (32\%)\\
    \setlength{\parskip}{1ex}\\
    4 – Year : Transfer & 976 (59\%) : 675 (41\%)\\
    \setlength{\parskip}{1ex}\\
    Full Time : Part Time & 1446 (88\%) : 205 (12\%)\\
    \setlength{\parskip}{1ex}\\
    IS : CS & 934 (57\%) : 717 (43\%)\\
    \setlength{\parskip}{1ex}\\
    1st Yr : 2nd Yr : 3rd Yr : 4th Yr & 115 (7\%) : 329 (20\%) : 515 (31\%) : 692 (42\%)\\
    \setlength{\parskip}{1ex}\\
    $\le$ 3 : 4-5 : $\gg$ 5  (Courses Enrolled) & 298 (18\%) : 1035 (63\%) : 318 (19\%)\\
    \setlength{\parskip}{1ex}\\
    \hline
  \end{tabular}
\end{table}

\section{Methodology}
\label{Methodology}

The methodology section of this study is divided into three subsections. In Section \ref{Method:feature extraction}, the focus is on the feature extraction methods that describes the methodology used to extract aggregated login related features from LMS data. Section \ref{Method: Correlation Studies} focuses on Correlation studies which comprises of machine learning approaches and correlation based model explanation methods deployed in the study. In section \ref{Method: causal-inference}, the methodology to study the causal relationships between student login behaviors, chronotypes, and academic performance outcomes are detailed.

\begin{figure}
    \centering
    \includegraphics[width=1.0\textwidth]{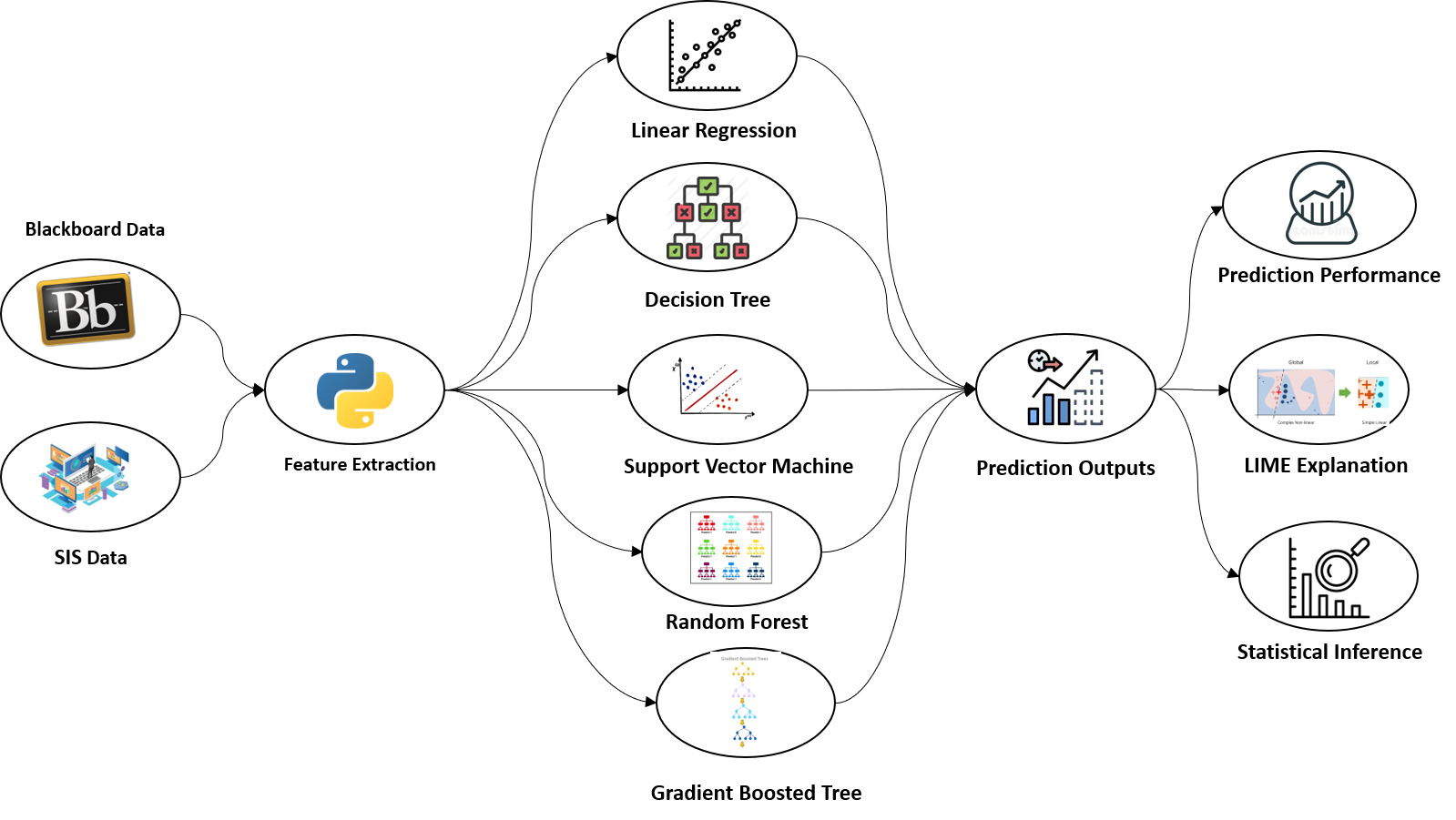}
    \caption{Student-centric Model Workflow}
    \label{fig:fig1}
\end{figure}

\subsection{Feature Extraction}
\label{Method:feature extraction}
From various LMS related features like student logins, content accesses, time spent, discussion posts, assignment submissions, and time intervals discussed in earlier literature, we identified that only three features could be commonly extracted from different courses to generate aggregate level student data: Student Login Counts, Time intervals and Prior performance.

One of the significant challenges while building a student-centric model on LMS data is to extract aggregated features that are least biased. As Blackboard's content is dependent on instructor and course, it is crucial to mitigate the variations caused by these factors on aggregate student variables. This work employs multiple statistical measures to mitigate these issues. The details are explained in the below Section \ref{Method: Normalized Login Volume}.

\subsubsection{Normalized Login Volume}
\label{Method: Normalized Login Volume}
Earlier studies showed that student performance is strongly correlated with the volume of student LMS logins. Also, calculating the total login count introduces a hidden bias as courses with more content on Blackboard prompt students to login more often than other courses with less content and flexible deadlines. To mitigate this issue, our work followed the below steps to extract student login features. 

\begin{enumerate}
    \item Extract all courses enrolled by the students in our study sample which includes students majored in both IS and CS;
    \item Count the total number of logins for all students enrolled in those courses, regardless if students are part of our study sample;
    \item Calculate normalized login volume for each student in our study sample in each course they enrolled. For a given student $S_i$ and a given course $C_j$ the student is enrolled in, the normalized login volume $N_{ij}$ is a Z-score calculated as $\frac{L_{ij} - \mu_j}{SD_j} $ where $L_{ij}$ is the absolute login counts for student $S_i$ in course $C_j$, and $\mu_j$ and $SD_j$ is the mean and standard deviation aggregated from all absolute login counts for all students enrolled in course $C_j$. 

    This step is crucial as it allows us to mitigate the bias from variations in courses design which is reflected in the between-course variations of overall login volume.  Z-scores provide a value that helps understand if student logins are higher or less than average logins in a specific course.
    \item Once the z-scores are calculated for all courses, we extract a vector of login z-scores for each student based on their enrolled courses.
    \item As most predictive models do not take vectors of variable length as input,  we then extracts seven significant statistics from the login vector: mean, median, minimum, maximum, standard deviation, skewness, and kurtosis.
    
\end{enumerate}

\subsubsection{Login Regularity}
\label{Method: Login Regularity}
Apart from student login volumes, the regularity of logins also provides valuable insights into student achievement as regularity is related to students' habitual study patterns which reflects their self-regulation capabilities. In this work, we utilize an \textit{entropy-based method} to extract features that define student login regularity in each course. In information theory, entropy is used to define uncertainty or randomness \cite{gray2011entropy}. Entropy measure will explain if student's logins are regular (less random) or irregular (more random). Based on this concept, if the entropy value is high, then a student has an irregular login pattern, and if the entropy value is low, the student has a regular login pattern. The steps to calculate student regularity features are given below.

\begin{enumerate}
    \item Extract all course accesses with timestamps for every student in our study sample.
    \item Calculate the interval between login timestamps which will result to a vector of time intervals for each course enrolled by a student.
    \item Calculate login interval entropy  $H_{ij}$ given a student $S_i$ and course $C_j$. Entropy is calculated using the KL estimator with the k nearest neighbor method proposed by Kozachenko and Leonenko \cite{kozachenko1987sample}. KL estimator uses k-nearest neighbor distances to compute the entropy of distributions. The reason for adopting this method instead of the classicial Shannon entropy is due to the continuous nature of time intervals \cite{kraskov2004estimating}.
    \item Once the entropies are calculated, we get a vector of entropies for each student based on the number of enrolled courses. We then calculate the seven statistics similar to student logins: mean, median, minimum, maximum, standard deviation, skewness, and kurtosis.
\end{enumerate}

\subsubsection{Login Chronotypes}
\label{Method: Login Chronotypes}
Studies in chronobiology and chronopsychology showed variation in different individual active periods at different times of the day \cite{putilov2021single,romo2020evening}. These studies classify an individual into either morning type or evening type based on their timing of high activity period. For example, if an individual is highly active in the morning compared to the evening, they are considered morning type and vice versa. Inspired by this work in human psychology \cite{putilov2021single}, this work divides a day into 24 hourly time bands and in addition aggregates 24 hours in a day into four-time bands T1 (12 AM to 6 AM), T2 (6 AM to 12 PM), T3 (12 PM to 6 PM), and T4 (6 PM to 12 AM) and extract student logins. In addition to this, this work also extracts the logins on weekdays and weekends to study their influence on student performance.

\begin{enumerate}
    \item Count the number of logins during each hour in a day, count number of logins based on 4 time bands T1-T4 and on weekdays and weekends for each course through out the semester.
    \item Calculate the mean of login count vector across all days in a given semester for each of these time bands and weekday/weekend.
    \item Normalize the login count with the number of courses enrolled by an individual student. This normalization will mitigate the bias introduced by the number of courses enrolled across the student cohort.
\end{enumerate}

\subsubsection{Student Demographics and GPA from prior semesters}
\label{Method: Student Demographics and GPA from prior semesters}
This work also utilizes the demographic and prior performance measured by GPA features captured by the SIS system. These student-level features were listed in Table~\ref{tab:2}

\begin{table}[!ht]
  \caption{Student demographics and GPA at beginning of semester.}\vspace*{1ex}
  \label{tab:2}
  \centering 
  \begin{tabular}{ c  c } 
  \hline
    \multicolumn{1}{c}{\textbf{Features}} & \multicolumn{1}{c}{\textbf{Values}} \\
    \hline
    \setlength{\parskip}{1ex}\\
    Start GPA (Prior Performance) & Cumulative GPA available till the start of semester \\
    \setlength{\parskip}{1ex}\\
    Gender & Male \& Female\\
    \setlength{\parskip}{1ex}\\
    Ethnicity & White, Asian \& Minority \\
    \setlength{\parskip}{1ex}\\
    Student Year & Freshman, Sophomore, Junior \& Senior \\
    \setlength{\parskip}{1ex}\\
    Admit Type  & Regular \& Transfer\\
    \setlength{\parskip}{1ex}\\
    Enrollment Type & Full time \& Part time \\
    \setlength{\parskip}{1ex}\\
    Student Age & Continuous variable\\
    \setlength{\parskip}{1ex}\\
    \hline
    \end{tabular}
\end{table}

\subsection{Correlation Studies}
\label{Method: Correlation Studies}
\subsubsection{Machine Learning approach}
\label{Method: Machine Learning approach}
This work studied five of the most common regression models for comparison purposes. The selected models include Regularized Linear Regression (LR), Decision Tree (DT), Support Vector Regressor (SVR), Random Forest (RF), and Gradient Boosted Regressor (GBR). As model hyperparameter influences their predictive performance, we utilized a grid search mechanism to select multiple parameters to predict with high accuracy. We also adopted a feature selection method based on a multi-objective evolutionary algorithm in addition to hyperparameter search. This feature selection algorithm evaluates each feature set based on pareto-optimal that balances model complexity and accuracy. The details of models and hyperparameter search criteria are discussed below.

Regularized Linear Regression: This model is extension from conventional linear regression. The regularization parameter is set so that the hyperparameter search space looks for an alpha value that fits between ridge and lasso regression. An alpha value of 1 represents lasso regression, and an alpha value of 0 represents ridge regression. This study searched for the best alpha value using a grid search between 0 and 1 in increments of 0.1.

Decision Tree: The decision tree algorithm is a collection of linked nodes intended to estimate the numerical target variable. Each node in the tree represents a rule used to split on an attribute value. The node uses a least-squares criterion to minimize the squared distance between the average value in a node when compared to the actual value. The hyperparameter search space for this algorithm evaluates both maximal depth and pruning. The maximal depth value varies between 1 and 100 in increments of 10. Pruning will make the DT algorithm use multiple criteria like minimal gain, minimal leaf size, and pruning alternatives to decide the stopping criterion.

Support Vector Machines: The SVM used in this study is built based on Stefan Reupping’s mySVM \cite{ruping2000mysvm}. This algorithm will construct a set of hyperplanes in a high dimensional space for regression tasks. A good hyperplane is decided based on the functional margin. The hyperparameter search space focused on both dot and radial kernel functions with a C (SVM complexity) value range between 10 and 200. The kernel gamma function is set for a radial kernel with a range of 0.005 and 5 with three logarithmic increments.

Random Forest: A RF model builds an ensemble of decision trees on bootstrapped datasets. The splitting criteria are similar to a decision tree. The regression outcome is the average of the observed train data GPA present at that end node. We only tuned the number of trees hyperparameter to reduce the time complexity of the execution. The number of tree searches varied between 10 and 1000 trees in 10 linear steps 

Gradient Boosted Tree: The GBT model builds multiple regression trees in a sequence by employing boosting method. By sequentially applying weak learners on incrementally changed data, the algorithm builds a series of decision trees that produce and an ensemble of weak regression models. As GBT is a nonlinear model, we search hyperparameters related to the number of trees, learning rate, and maximal depth. The number of tree values varies between 1 and 1000 in five quadratic increments, the learning rate varies between 0.001 and 0.01 in five logarithmic increments, and the maximal depth parameter varies between 3 and 15 in three logarithmic increments

\subsubsection{Model Explanation using LIME}
\label{Method: Model Explanation using LIME} 
The concept of Locally Interpretable Model Explanations (LIME) was introduced to explain the predictions made by black-box models that deal with classification problems. LIME explains each prediction made by a complex model by training a surrogate model locally \cite{ribeiro2016should}. However, this earlier methodology is not scalable to deal with categorical variables, tabular data, and regression problems. In this work, we adopt the correlation-based LIME method available in RapidMiner \cite{rapidminer} to explain machine learning models' predictions\cite{hofmann2016rapidminer,mandalapu2019studying,mandalapu2021we,mandalapu2021profiling}. The steps invovled in LIME is as following:

\begin{enumerate}
    \item Perturb data in the neighborhood of each sample in the dataset. The number of simulated samples can be defined by user. A higher number of simulated samples will provide higher accuracy of explanations but at the cost of more computing time.
    \item Make predictions using the ML model for all the simulated samples around each original sample in the dataset.
    \item Calculate the correlation between each feature in the dataset and the target variable.
    \item The features that have a positive correlation are considered supporting features, and features with negative correlation with predicted outputs are referred to as contradicting features.
\end{enumerate}

As LIME provides feature importance value for each feature at each sample, we aggregate the importance value for all samples to build global importance for each variable. The significant advantage of this method compared to traditional global importance methods is its flexibility. As model global importance’s are calculated across all samples in the data, the LIME based feature importance’s can be calculated for subsets of data. This flexibility provides users with a deeper understanding of each feature's role for different sets of populations present in a dataset.

In addition to applying the LIME methodology, this work also studies univariate and multivariate feature importance on student performances by applying correlation and linear regression methods. The student dataset used in this study is divided into multiple subsets containing different student groups based on various demographics. A correlation value is calculated between input features and student end-of-term GPA. This value provides us with an intuition about the impact of various features on student performances related to different demographics. As correlation only provides independent variable importance on student performance, we also adopt a linear regression model to explore the variation of feature importance based on coefficient values. Applying a linear regression model will also consider the interaction effect between input features to fit the outcome variable.

\subsection{Causal Inference}
\label{Method: causal-inference}

\begin{figure}[h!]
    \centering
    \includegraphics[width=1.0\textwidth]{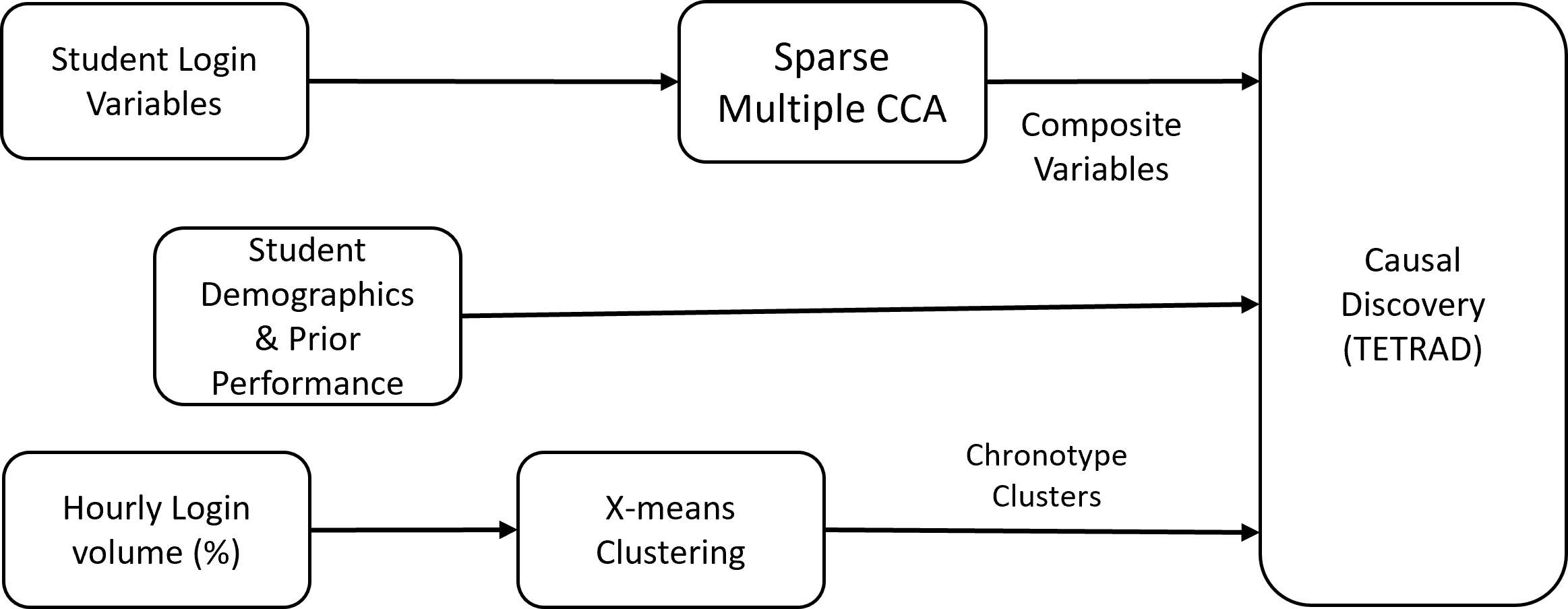}
    \caption{ Analysis pipeline to study causal relationships between student learning behaviors, chronotypes and performance}
    \label{fig:fig2}
\end{figure}

One of the primary objectives of this study is to understand any causal relationships between student login behaviors and their performance outcomes. This understanding will support the development of efficient and effective intervention strategies that may positively impact student academic performance, effectively turning data into insights and into action. To achieve this goal, we adopt the processing pipeline shown in Figure \ref{fig:fig2}. 

The process shown in Figure \ref{fig:fig2} takes three inputs: student learning behaviors as characterized by login variables as described in Section \ref{Method:feature extraction} , demographics, and chronotype clusters. For login variables, we specifically focus on normalized student login volume, regularity of student logins, and weekday/weekend logins described in Table \ref{tab:3}. As these features are a group of independent statistics (mean, median, standard deviation, minimum, maximum, skewness, and kurtosis) grouped into sets based on their relevance, we input them to Sparse Multiple Canonical Correlation Analysis (CCA) algorithm. This algorithm will help learn sparse representation for each feature group by maximizing the correlation among these feature sets with multiple features \cite{hardoon2011sparse}. This CCA approach results in a composite variable for each feature group. These are represented by a linear combination of a small set of features and are used as inputs for causal structural discovery and inference.

Student demographics are used in two ways. In the first method, we feed these demographics directly to the causal discovery model to study their relationships with student performances. In the second method, we use this demographic information to group students based on their demographics and explore causal relationships for students belonging to a specific group. Finally, we use chronotype clusters to study the causal relationship between chronotypes and student performance.

\subsubsection{Causal Inference and Discovery}
\label{Method: Causal Inference Discovery}
This work's causal discovery and inference are performed using a software suite named TETRAD \cite{ramsey2018tetrad}. TETRAD was developed 20 years ago as a drag and drop suite of procedures to explore causal relationships in the input dataset. This suite can take continuous, categorical, mixed, covariance, and correlations as input for causal discovery. This tool consists of multiple proven algorithms that are selected based on the type of data that is inputted for causal discovery. This work adopts two algorithms: Greedy Fast Causal Inference (GFCI) and PC algorithm \cite{ogarrio2016hybrid,spirtes1991algorithm}. These algorithms were selected based on their ability to accept prior knowledge that is useful to control the directionality of causal relations.

The GFCI algorithm only works with continuous variables and can output Partial Ancestral Graphs (PAG).  These PAG's are causal networks that include hidden confounder variables. The working of this algorithm is based on the combination of two algorithms named FGES and FCI \cite{spirtes1993discovery}. The FGES algorithm takes input sample data and background knowledge to score using a greedy search mechanism and applies it to a larger sample, and selects the Causal Bayesian Network (CBN) with a higher score. On the other hand, FCI is a constraint-based algorithm with similar functionality to FGES. In addition to this, it entails the set of conditional independence relations that will be satisfied at population-level data. FCI algorithm has two phases named as adjacency phase and orientation phase. The adjacency phase starts with an undirected graph and performs a sequence of conditional independence tests to filter edges between two adjacent variables that are independent. As the output of the adjacency phase is an undirected graph, the directionality between adjacent sets is provided by the orientation phase. In the orientation phase, the stored conditional settings used to remove the adjacencies to reduce edges are used to determine directionality. GFCI algorithm uses FGES to improve both adjacency and orientation phases of FCI by providing a more accurate initial graph for causal network development. 

Though GFCI is a robust causal inference algorithm for continuous datasets, it is unable to handle categorical variables like student demographics. As such, we only used the GFCI algorithm to study the causal relationship between login variables and student performance.  

We adopt a PC algorithm to explore causal relationships with both categorical and continuous variables to mitigate this issue. PC algorithms utilize a pattern search method that assumes the underlying structures in data are acyclic. This algorithm also assumes that no two variables have the same latent variable. The only drawback of the PC search algorithm is its inability to show confounding relations in its data, whereas GFCI algorithm mitigates this issue. This is also one of the primary reasons to adopt two algorithms in this study instead of one for causal structure discovery and inference. 

\subsubsection{Sparse Multiple CCA}
\label{Method: Sparse Multiple CCA}

The high dimensional dataset in this study is reorganized into multiple sets based on their roles in student LMS interaction process. The primary reason for reorganizing features variables into different groups is to make the causal inference part of this study more interpretable. The variables are groups with meaningful roles is shown in Table \ref{tab:3}. But reorganizing variables into different groups alone is not sufficient for interpretability as causal algorithm needs a single variable to represent each feature concept related to LMS interactions. We generates a composite variable from the group of features for every feature subsets using Sparse Multiple Canonical Correlation Analysis (mCCA) \cite{hardoon2011sparse}. The traditional CCA method takes two matrices as inputs to generate a linear combination of variables in each feature set with a high correlation between the two feature sets. This is similar to the way Principal Component Analysis (PCA) works, but on multiple datasets at a time. In addition to this, CCA also maximizes the correlation between datasets while generating composite variables. Sparse CCA is an extension to traditional CCA, where sparsity constraints are imposed. This makes it more compact and yields a more interpretable representation of the data. This study adopts the mCCA function from the PMA package in R. The standardized student login variables were inputted to the mCCA algorithm. A grid search is performed to select the model with the best total correlation and compactness. This model is used to output weights for each feature and then generate composite variables for each feature set. These composite variables are then fed to GFCI or PC variant algorithm to perform causal structural discovery and inference.

\begin{table}[ht!]
  \caption{Feature set for causal analysis relationships.}\vspace*{1ex}
  \label{tab:3}
  \centering 
  \begin{tabular}{ c  c } 
  \hline
    \multicolumn{1}{c}{\textbf{Feature}} & \multicolumn{1}{c}{\textbf{Description}} \\
    \hline
    &  Age, Gender, Student Year, Type of Admit, \\
    Demographics & Full time/Part time, Student Ethnicity.\\\hline 
    & The total number of courses enrolled by \\
    Enrolled Courses & student in Fall 2019\\\hline
    & The cumulative GPA of student till the \\
    Student Start GPA & start of Fall 2019 Semester\\\hline
    & Statistics related to login z scores of \\
    & students per each enrolled course: Mean, \\
    Normalized Login Volume & Median, Standard Deviation, Minimum Score,\\
    & Maximum Score, Skewness, and Kurtosis.\\\hline
    & Statistics from KL Entropy of student \\
    & login intervals per each enrolled course:\\
    Student Regularity in Logins & Mean, Median, Standard Deviation,\\
    & Minimum Score, Maximum Score, Skewness, and Kurtosis.\\\hline
    Weekday and Weekend volumes & Login counts on weekdays and weekends.\\\hline
    & Count cumulative login volumes per hour \\
    Hourly Login Volume Percentage & and normalize them by calculating the\\
    & percentage of logins per hour.\\\hline
    & Count the aggregate time spent per 60\\
    & minutes every hour till the middle of the\\
    Hourly Time spent & semester and then calculate the\\
    & percentages by normalizing across 24-hour bands in a day.\\
    \hline
  \end{tabular}
\end{table}

\section{Results}
\label{Results}
In this section 6, we discuss the outcomes of experiments performed in this study. Section \ref{Result: Chronotypes Analysis} reports the results of clustering student activity at different times in a day, referred to as chronotype analysis. In the section \ref{Result: Correlation Analysis}, we report the performance of five machine learning algorithms to predict student performance at the end of semester by learning from student login, interactions, and demographic data till the middle semester. This subsection also reports the importance of different features based on the correlation between predictors and target variable and also using LIME post hoc model explanation. The final section \ref{Result: Causal Analysis} reports the outcomes of causal inference and discovery experiments to study the cause and effect between different predictor variables and also between predictor variables and outcome variable.

\subsection{Chronotypes Analysis}
\label{Result: Chronotypes Analysis}
\begin{figure}
    \centering
    \includegraphics{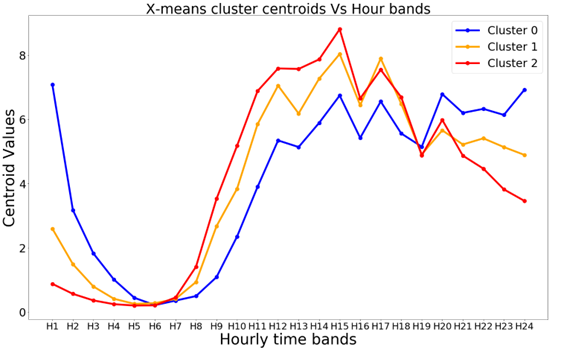}
    \caption{Clustering outcome of normalized hourly login volume data to study student chronotype}
    \label{fig:fig3}
\end{figure}

We cluster the student hourly login percentages using X-means with Dynamic Time Warping (DTW) \cite{muller2007dynamic} distance calculation to study different chronotypes. The results of the clustering algorithm are shown in Figure \ref{fig:fig3}. As shown, the X-means algorithm found three clusters to be optimal for the input data. The x-axis in Figure \ref{fig:fig3} shows the hours in a day based on the 24-hour clock. It starts with H1 representing 12 AM to 1 AM and ends with H24 describing 11 PM to 12 AM with 1-hour incremental time bands. The y-axis in Figure \ref{fig:fig3} represents cluster centroids for each corresponding time band and cluster.

From cluster 0, we can observe that students' login percentages slowly start to increase from the morning at 8 AM (H9) and reaches a peak between 12 AM and 1 AM (H1). We can classify these students as most active in the evening with peak activity at night. There are a total of 239 students clustered into this group. Cluster 1 in Figure \ref{fig:fig3} starts to see an activity increase from 8 AM and reaches a peak activity at 2 PM and 4 PM and suddenly drops till 7 PM in the evening and gradually stabilizes there till 11 PM and then dips during the nighttime. There is a total of 560 students in cluster 1 that are assigned to this pattern. In final cluster 2, consisting of 889 students, the pattern is similar to cluster 1, but it doesn't stabilize at 7 PM. The dip in activity of these students continues with very few of them active at night. Based on the patterns found in these three clusters, we can classify cluster 0 as "Afternoon to Night," cluster 1 as "Afternoon to Evening," and cluster 2 as "Active Afternoon." Even though these three clusters have some similarities in the afternoon login patterns, where most are active, we can also see a peak in the evening between 7 PM and 8 PM. This specific time slot is interesting as the variations in cluster patterns started in the evening except for this particular time band. To understand any relationship between student demographics and chronotype clusters, we adopt the chi-square significant test and record the p-values that are listed in Table \ref{tab:4}. From Table \ref{tab:4}, we can observe a meaningful relationship between student ethnicity and chronotypes. Another significant relation is found between student enrollment type (full-time or part-time) and their chronotypes. This can be related to the varying course loads and a difference in the nonacademic workload of part-time students compared to full-time students. Finally, as a student progresses through their undergraduate years, we can see a shift in chronotypes. 

\begin{table}[h]
  \caption{Student Demographics and their corresponding p-values based on chi-square test with Chronotype Clusters.}\vspace*{1ex}
  \label{tab:4}
  \centering
  \begin{tabular}{ c  c }
    \hline
    \multicolumn{1}{c}{\textbf{Demographic}} & \multicolumn{1}{c}{\textbf{P-value}} \\
    \hline
    Gender & 0.0571\\
    \textbf{Ethnicity} & 0.0001\\
    \textbf{Full Time or Part Time} & 0.007\\
    Regular or Transfer & 0.384\\
    \textbf{Student Year} & 0.018\\
    GPA Bands & 0.855\\
    \hline
  \end{tabular}
\end{table}

\subsection{Correlation Analysis}
\label{Result: Correlation Analysis}

\subsubsection{Predictive Model Performance}
\label{Result: Predictive Model Performance}

The five machine learning models adopted in this study were evaluated using a five-fold cross-validation method. In this method, the student data is divided into five equal folds at a student level. Four of the five folds are used for model training in every iteration, and one fold is used for model testing. The machine learning models are evaluated based on two performance metrics: R squared $(R^{2})$ and Root Mean Squared Error (RMSE). The output performance metrics are the average of five test fold performances. From the table \ref{tab:5} we can see that both Random Forest and Gradient Boosting Tree performed similarly with a R-squared value of 0.36 and RMSE value of 0.68.

\begin{table}
  \caption{Performance evaluation of predictive models based on R squared and RMSE values.}\vspace*{1ex}
  \label{tab:5}
  \centering
  \begin{tabular}{ c  c  c}
    \hline
    \multicolumn{1}{c}{Model} & \multicolumn{1}{c}{R Squared} &  \multicolumn{1}{c}{RMSE}\\
    \hline
    LR & 0.26 & 0.73\\
    DT & 0.24 & 0.74\\
    SVM & 0.20 & 0.77\\
    RF & 0.36 & 0.68\\
    GBT & 0.36 & 0.68\\
    \hline
  \end{tabular}
\end{table}

We also validated the best performing model (GBT) resulted from current and earlier study on Fall 2019 middle of semester data by testing them on Spring 2019 and Fall 2018 student data related among IS and CS students. The results in Table \ref{tab:6} shows that student-centric models can be generalized across different semesters. Additionally, this work tested further the the model's external validity by validating them on student data from other departments in the Fall 2019 semester. From Table \ref{tab:7}, we can observe that the test performance of the model developed on IS and CS student data showed similar performance metrics on students enrolled in four other departments.

\begin{table}
  \caption{Validating the generalizability of models developed on Fall 2019 student data to different terms.}\vspace*{1ex}
  \label{tab:6}
  \centering
  \begin{tabular}{ c  c  c  c}
    \hline
    \multicolumn{1}{c}{Term} & \multicolumn{1}{c}{R-squared} &  \multicolumn{1}{c}{RMSE} & \multicolumn{1}{c}{Student Count (IS \& CS)}\\
    \hline
    Fall 2019 (Main Model) & 0.362 & 0.622 & 1559\\
    Spring 2019 & 0.370 & 0.683 & 1545\\
    Fall 2018 & 0.360 & 0.722 & 1486\\
    \hline
  \end{tabular}
\end{table}

\begin{table}
  \caption{Validating the applicability of models developed on IS \& CS student data to students from other departments}\vspace*{1ex}
  \label{tab:7}
  \centering
  \begin{tabular}{ c  c  c  c c}
    \hline
    \multicolumn{1}{c}{Department} & \multicolumn{1}{c}{Degree} &  \multicolumn{1}{c}{R-squared} & \multicolumn{1}{c}{RMSE}  & \multicolumn{1}{c}{Student Count}\\
    \hline
    IS \& CS (Main Model) & BS & 0.362 & 0.622 & 1559\\
    Biological Science & BS & 0.381 & 0.675 & 491\\
    Bio Chemical \& Molecular Biology & BS & 0.381 & 0.737 & 226\\
    Mechanical Engineering & BS & 0.385 & 0.657 & 391\\
    Psychology & BS & 0.317 & 0.849 & 215\\
    \hline
  \end{tabular}
\end{table}

\subsubsection{Model Explanation}
\label{Result: Model Explanation}

Given the large set of features we are exploring, we are interested in gaining insights of the relative importance of features in predicting students' performance. This step is crucial when we move the model from the lab into the real-world to inform decision-making.  We explored three different approaches that can provide complementary views into the model insights. While LIME method attempts to explain blackbox models, simple and linear model based approach interpretable by design. In explaining the model, we pay special attention to the role of students' demographics attributes may play in revealing the heterogeneity of feature importance in predicting student outcomes.



\paragraph{\textbf{Approach \#1: LIME based Feature Importance}}

LIME-based approach extract feature importance at the local level also called local fidelity. By applying the LIME method explained in the methodology section, we extract feature importance for different student groups categorized based on their demographics.

\begin{figure}
    \centering
    \includegraphics[width=0.8\textwidth]{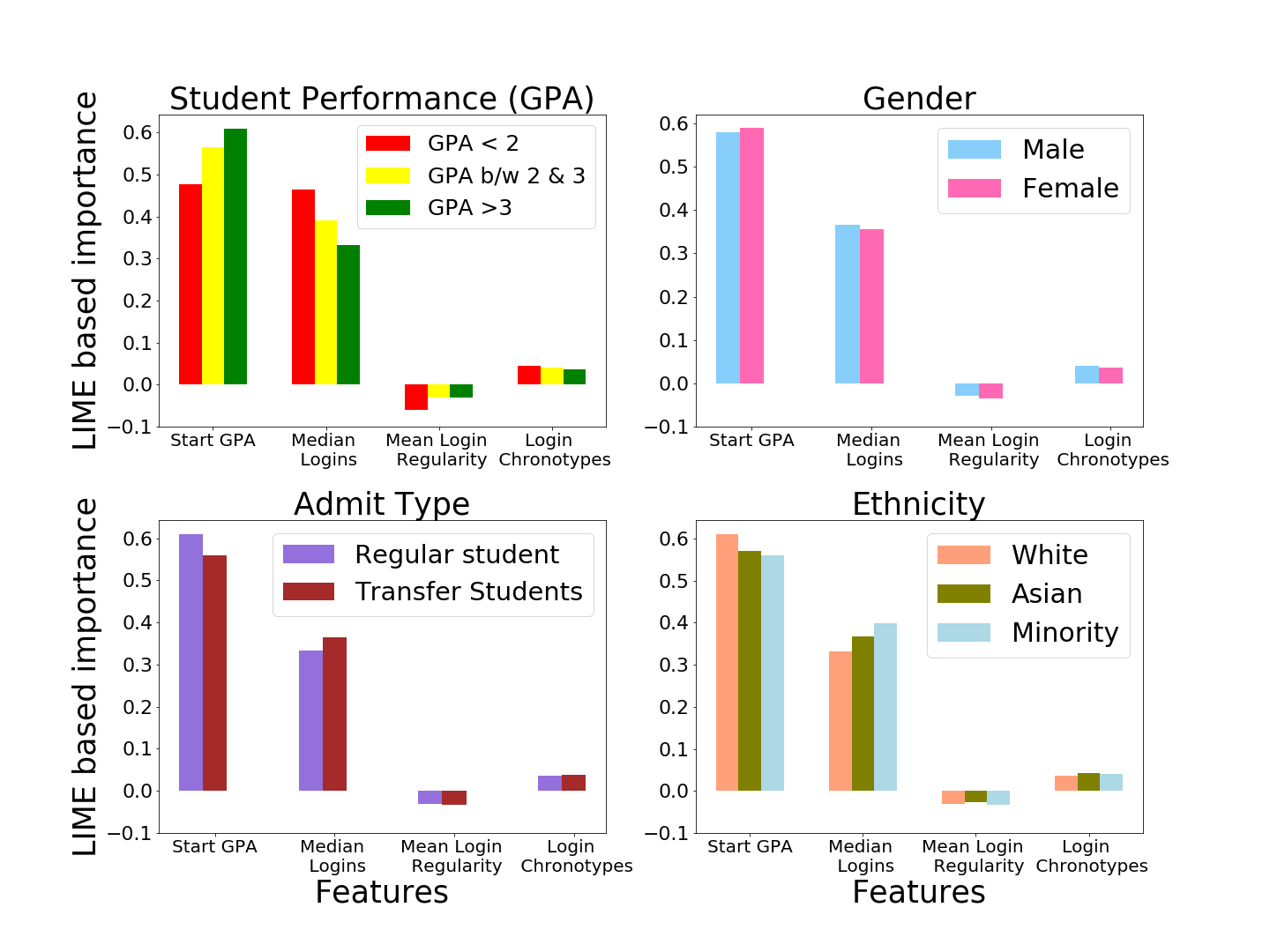}
    \caption{LIME importance’s for different student groups divided based on GPA, ethnicity, admit type and gender.}
    \label{fig:fig4}
\end{figure}

From Figure~\ref{fig:fig4}, we can observe that cumulative student GPA at the start of the semester is an important feature to predict student end-of-term GPA. Student login volumes are the second important feature set for model predictions on different student demographics. This study's focus is also on student self-regulation capability measured by the regularity of logins (entropy).  This entropy measure is labeled as "mean login regularity" in \ref{fig:fig4} and in general higher entropy means less regular. We observe that for students with GPA values less than 2, the regularity of the logins feature played a key role compared to a student with a higher GPA. In case of minority students, we observe that login volumes have higher importance on their performance when compared to other student ethnic groups. We also observe that the importance of login regularity to predict academic performance does not have significant variation when compared between different ethnic groups. One implication from these observations is that introducing teaching practices that guide LMS use and time management will significantly impact students with low GPA and a minority race. Start GPA played a slightly less significant role in transfer students than regular students as transfer students join in different years. Their cumulative GPA might not be available at the start of the semester, similar to freshman.

Even though there is a huge imbalance in the number of male and female students present in the dataset, we do not observe any significant difference in feature importance between these two genders. One limitation of the LIME method is related to global importance. The importance showed by LIME at the local level do not necessarily correspond to global importance.

\paragraph{\textbf{Approach \#2: Correlation-based Feature Importance}}

As earlier LIME based feature importance methods showed a significant impact of login volumes to predict student performance, we adopt Pearson correlation statistic to infer this relationship for different student groups. To do this, we create subsets of student data based on different groups: student GPA, gender, ethnicity, and admit type.

\begin{figure}
    \centering
    \includegraphics[width=0.8\textwidth]{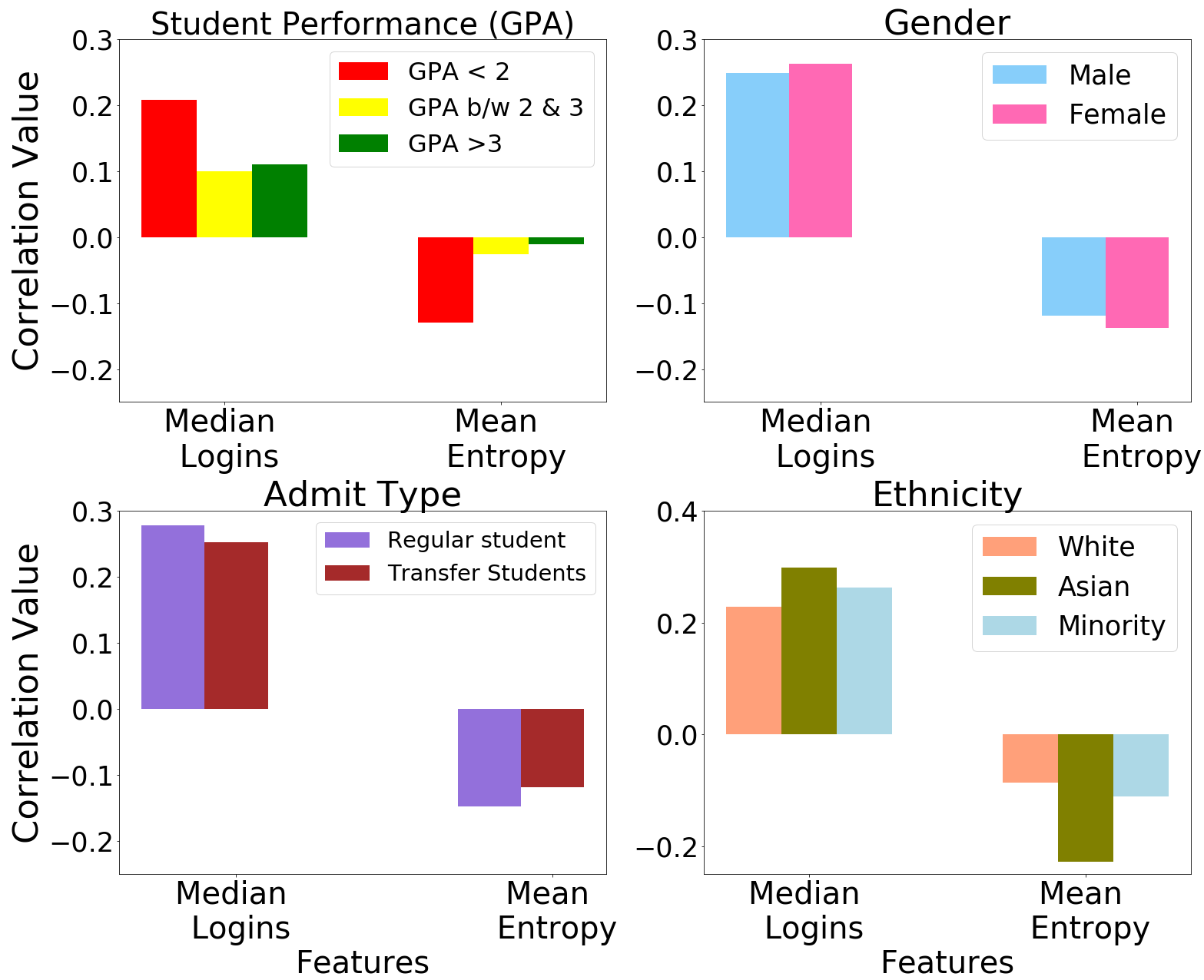}
    \caption{Correlation values for different student groups divided based on GPA, ethnicity, admit type and gender.}
    \label{fig:fig5}
\end{figure}

From Figure~\ref{fig:fig5}, we observe that the student login count and regularity in logins are highly significant for a student with a GPA lower than 2. We can also observe that as the entropy increases (i.e. students following a less regular schedule), the GPA reduces, as inferred from the negative correlation between the two variables. This observation seems to corroborate the understanding that regularity in student logins reflects their higher self-regulation capabilities which correlate with better learning outcomes. Earlier research showed that students with good self-regulation capabilities perform better in class \cite{kim2014predicting,park2015development}. For other student groups divided based on gender and admit type, there is no significant variation in the importance of login volume and login regularity on student performances.

\paragraph{\textbf{Approach \#3: Regression Model Based Feature Importance}}

One significant limitation of earlier methods is their inability to capture interaction effects as feature importance might change in the presence of other features. To study the interaction effects, we apply a linear regression model on different categories of student login data collected till the middle of the semester. These student categories were divided based on GPA, gender, admit type, and ethnicity of students. Even though linear regression models are applied on all features discussed in earlier sections, we only report the coefficients of median login volume and mean login regularity in Table~\ref{tab:8}, as these variables are the focus of this study. From Table~\ref{tab:8}, we observe that login volume and login regularity features follow similar directions for students with lower GPAs and students from minority ethnic backgrounds as observed in the LIME and correlation-based analysis. Another reason for focusing on a student from these two groups is their higher attrition rates found in earlier studies \cite{kupczynski2011impact,liu2011high}. At least at surface level these findings suggest a potential benefit for targeting these two student groups as these groups of students happens to have higher attrition rates.

\begin{table}
  \caption{Regression coefficients (Significance marked with *)}\vspace*{1ex}
  \label{tab:8}
  \centering
  \begin{tabular}{p{2cm} p{2cm} p{2cm} p{2cm}}
    \hline
    \multicolumn{1}{p{3cm}}{Student Demographic} & \multicolumn{1}{p{3cm}}{Student Groups} &  \multicolumn{1}{p{3cm}}{Median Logins Coefficient} & \multicolumn{1}{p{3cm}}{Mean Login Regularity Coefficient}\\
    \hline
     \multirow{3}{2em}{GPA} & GPA $\le2$ & \textbf{0.171*} & \textbf{-0.398*}\\
     & GPA \textgreater2 \& $\le3$ & -0.013 & 0.200\\
     & GPA \textgreater3 & 0.065 & -0.002\\
     \setlength{\parskip}{1ex}\\
      \multirow{2}{2em}{Gender} & Male & 0.135 & 0.130\\
      & Female & -0.021 & -0.157\\
      \setlength{\parskip}{1ex}\\
       \multirow{2}{2em}{Admit Type} & Regular & 0.399 & -0.004\\
       & Transfer & 0.611 & 0.191\\
       \setlength{\parskip}{1ex}\\
        \multirow{3}{2em}{Ethnicity} & White & 0.204 & -0.029\\
         & Asian & -0.085 & 0.201\\
        & \textbf{Minority Race} & \textbf{0.201*} & \textbf{-0.153*}\\
    \hline
  \end{tabular}
\end{table}

\subsection{Causal Analysis}
\label{Result: Causal Analysis}

\subsubsection{Aggregated Level Causal Analysis}
\label{Result: Aggregated Level Causal Analysis}
This part of this study explores the causal relationships between student login variables, chronotypes, and performance outcomes treating students from a homogeneous group. We adopt PC variant and GFCI causal discovery algorithms. As discussed in \ref{Method: Causal Inference Discovery} the PC variant algorithm gives high level understanding of variables and is able to incorporate categorical variables whereas GFCI algorithm can indicate confounding relationships and can be used to quantify relationships based on Structural Equation Modeling. Our focus is to study how login variables are interacting with performance and the impact of demographics incase of PC variant algorithm and studying indepth quantifiable interaction between login behaviors and student performance with GFCI algorithm. To do this, we apply a PC Variant algorithm in TETRAD that can handle both categorical and continuous variables. The composite variable input for the PC algorithm is calculated based on mCCA. In addition, we also incorporate common-sense knowledge to constraint the directionality of cause and effect. For example, we stipulate that demographic variables can only be a cause not an effect as they cannot be changed. The login variables are both cause and effect as they are malleable with interventions or behavioral changes. These variables act as a causal explanation for end-of-term GPA but can be affected by student demographics. 

\begin{figure}[h!]
    \centering
    \includegraphics{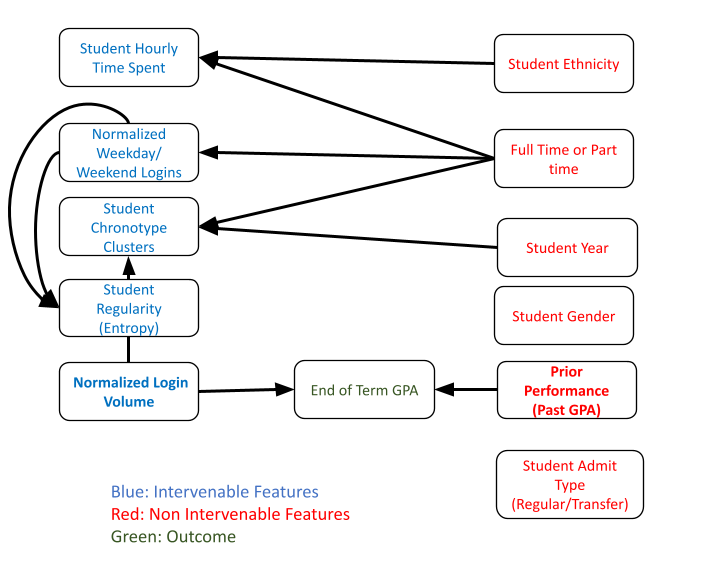}
    \caption{Causal relationship exploration between student demographics, login behaviors and chronotypes using PC Variant Algorithm.}
    \label{fig:fig6}
\end{figure}

The feature weights related to mCCA are listed in Appendix Table \ref{tab:D.1}. The causal relationships between variables are shown in Figure~\ref{fig:fig6}. From Figure~\ref{fig:fig6}, we observe that the student academic performance has a direct causal relationship with Normalized login volume and student prior performance. In addition, this analysis also shows that chronotype is causally linked to student enrollment type (part-time or full-time), and student year. Other demographics like gender and admit type (regular/transfer students) does not causally linked to login variables nor performance. Even though there are other causal relationships between different student variables, they are not directly impacting the performance. All the relationships displayed in Figure~\ref{fig:fig6} are significant as we set a p-value cutoff at 0.05.

One of the limitation of the PC variant algorithm is the lack of confounder identification. To identify if the causal relations are more robust or if there is a presence of any confounder, we employ the GFCI algorithm in TETRAD. Additionally, we also explore the causal relationships by dividing student data into subsets based on their different demographic features. The interpretation of causal connectors related to GFCI is given in Table~\ref{tab:9}.

\begin{table}
  \caption{Graph connector type descriptions}\vspace*{1ex}
  \label{tab:9}
  \centering
  \includegraphics{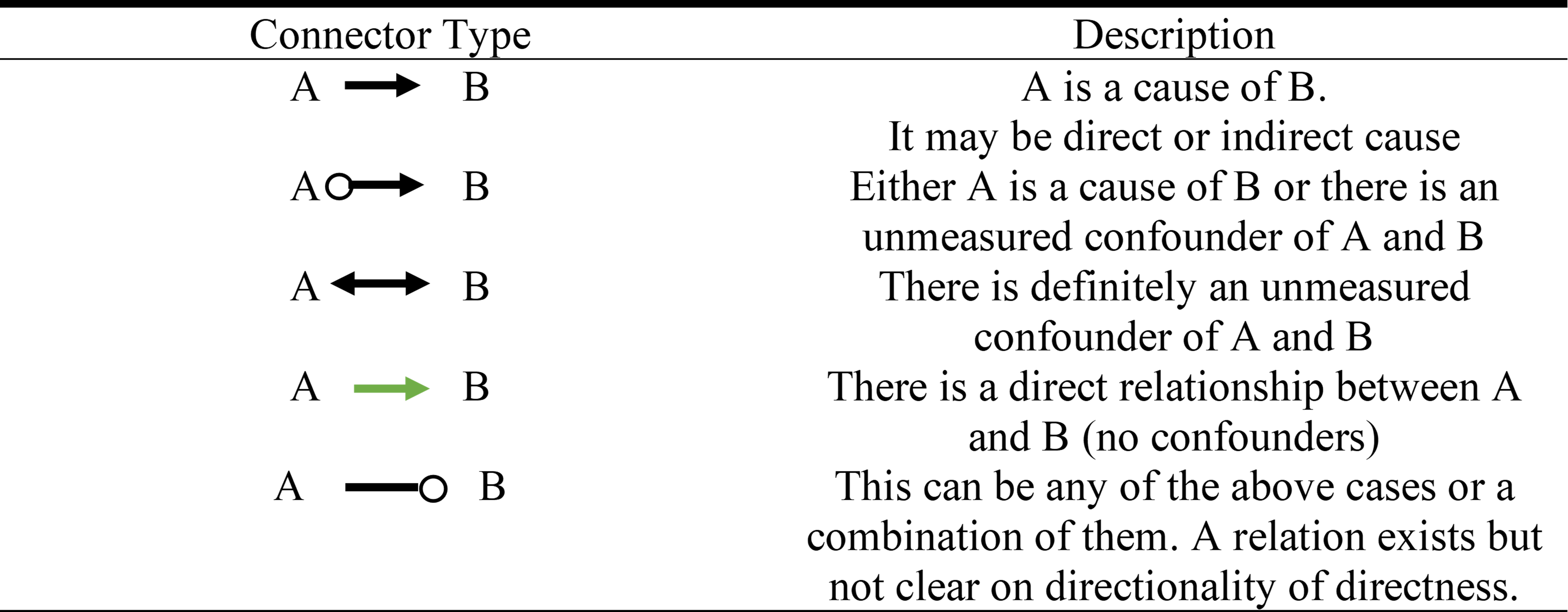}
\end{table}

\begin{figure}
    \centering
    \includegraphics{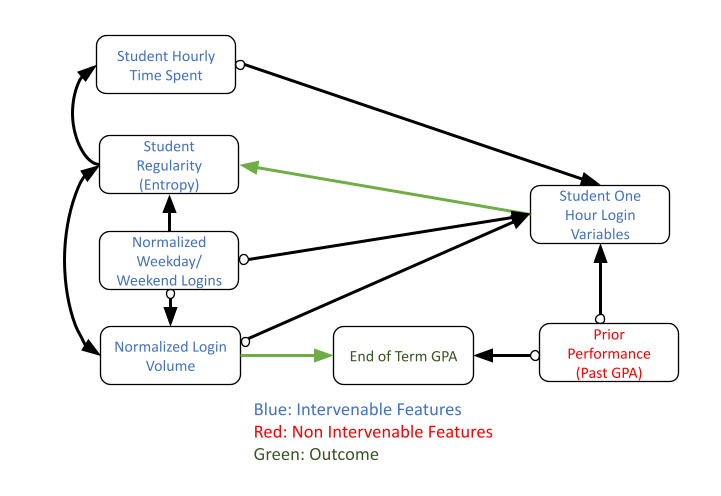}
    \caption{Causal relationship between Student login variables and their academic performance using GFCI algorithm}
    \label{fig:fig7}
\end{figure}

Figure~\ref{fig:fig7} shows the output of the GFCI algorithm related to continuous student login variables. It shows a direct causal relationship between student normalized login volume and their academic performance from this Figure~\ref{fig:fig7}. The relationship between student prior performance (Cumulative GPA till the start of the semester) and the current end of term GPA observed in PC variant analysis shown in Figure~\ref{fig:fig6} also exists in GFCI output. But it is not clear if this is a direct relationship or if there might be an unmeasured confounder between these two. Consistent relations identified by PC Variant and GFCI methods are found between student login regularity \& student login volume, student hourly login volume \& regularity, and Weekend/WeekDay login volumes \& login volumes as well as regularity. Even though multiple other relationships are common in both PC Variant analysis and GFCI for different variables, we can see no direct relationships that impact student performance except login volume and prior performance.

It should be noted that Figure~\ref{fig:fig7} does not fully specify causal relationships as there are some relationships with confounding factors. In addition to this, the GFCI causal graph in Figure~\ref{fig:fig7} only shows which variable acted as cause and which variable had an effect but it does not detail the strength and directionality of cause and effect relationship. The causal graph shown in Figure~\ref{fig:fig7} is used to generate a Directed Acyclic Graph (DAG) based on domain knowledge. This DAG is used as an input to estimate linear Gaussian Structural Equation Models (SEM). We then calculate the model's goodness of fit and coefficients to study if the causal relationship between the features is as expected based on domain knowledge. For example, for the pair of variables prior performance and End of Term GPA, we believe that there is a direct relationship between these two as prior performance can only be a cause but not an effect. For other variables with confounding measures, we can expect any directionality of cause and effect relationship as they all are relevant and drawn from the similar statistic. This is the reason we assign a double-headed arrow to other relationships with confounding factors.

\begin{figure}[ht]
    \centering
    \includegraphics{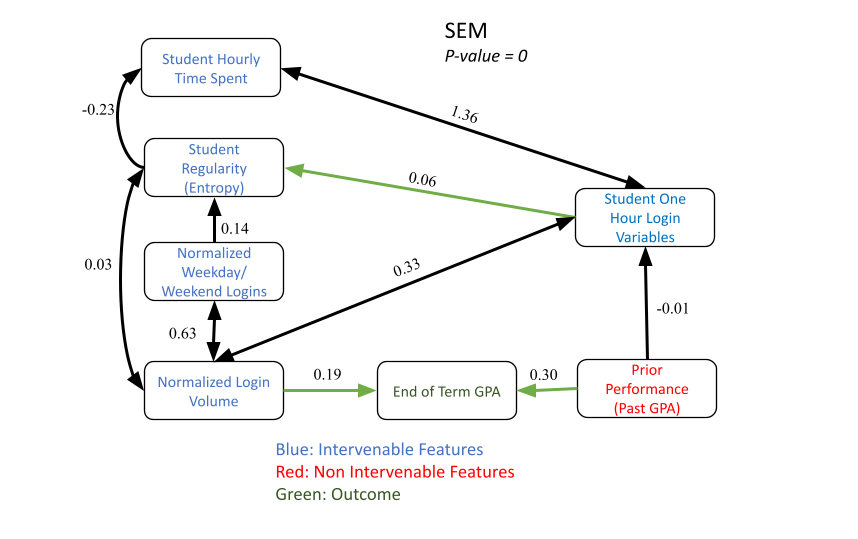}
    \caption{DAG derived from GFCI causal graph. Numerical values on directed arrows represent coefficients, and the values bidirectional arrows represent covariance reported by SEM}
    \label{fig:fig8}
\end{figure}

Figure~\ref{fig:fig8} represents DAG derived from the causal graph in Figure \ref{fig:fig7}. The values indicated on arrows represent coefficients if it is a unidirectional arrow and covariance if it is a bidirectional arrow. From this DAG, we can observe that Login volume causes End of term GPA and if the login volume increases, student GPA increases as the coefficient value (0.19) is positive. A similar relationship can be observed between student prior performance and current term end GPA. From Figure~\ref{fig:fig8}, we can confirm that the causal relationships observed in the GFCI algorithm are appropriate in the given dataset but doesnt scale to population as p value is close to 0.

\subsubsection{Group specific causal analysis}
\label{Result: Group specific causal analysis}
\begin{figure}
    \centering
    \subfigure[]{\includegraphics[width=0.9\textwidth]{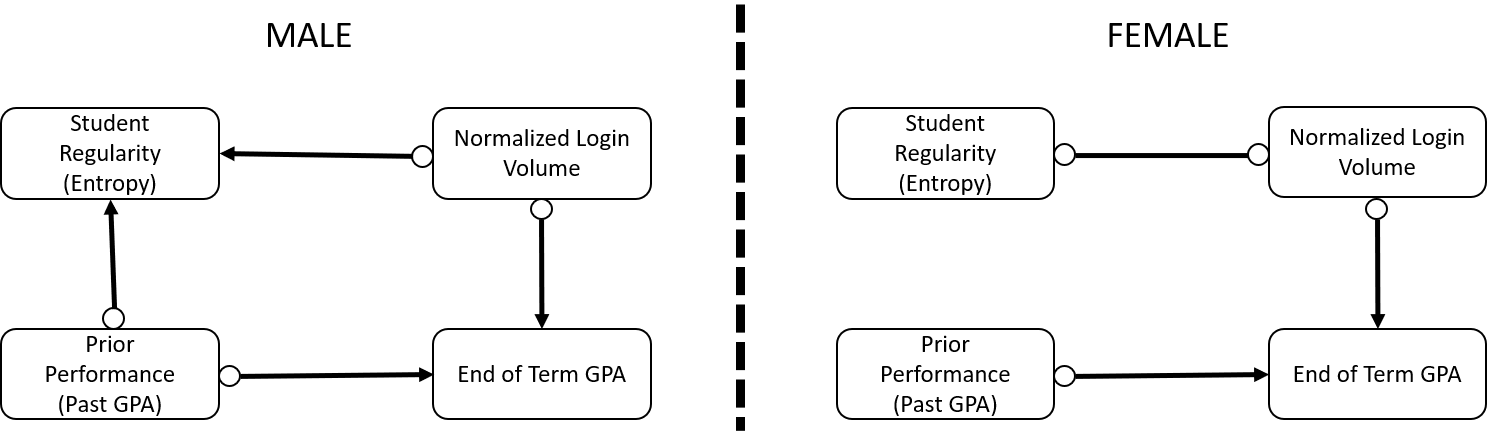}}
    \subfigure[]{\includegraphics[width=0.9\textwidth]{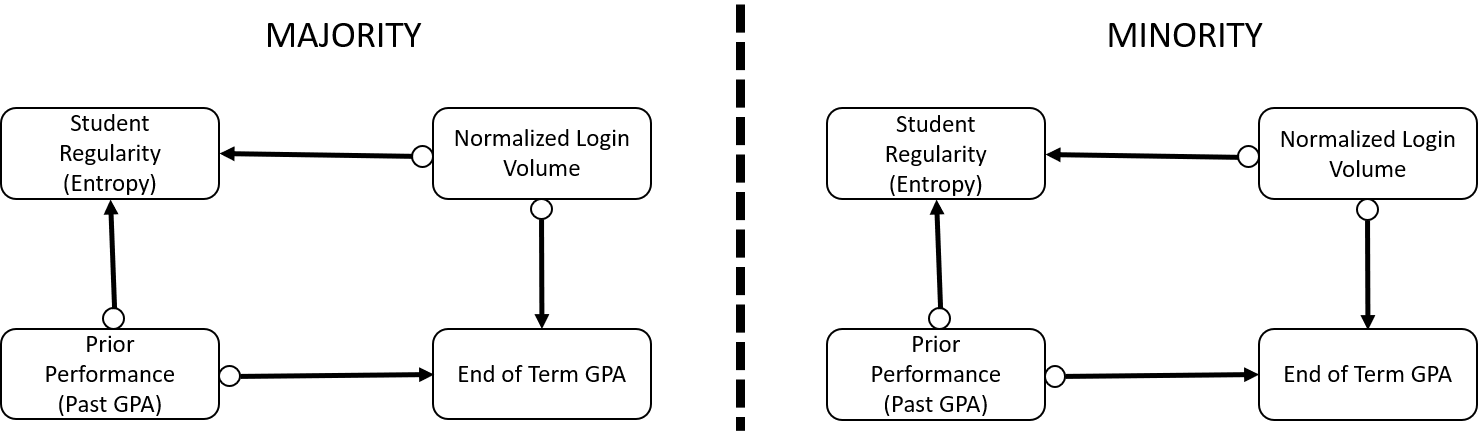}} 
    \subfigure[]{\includegraphics[width=0.9\textwidth]{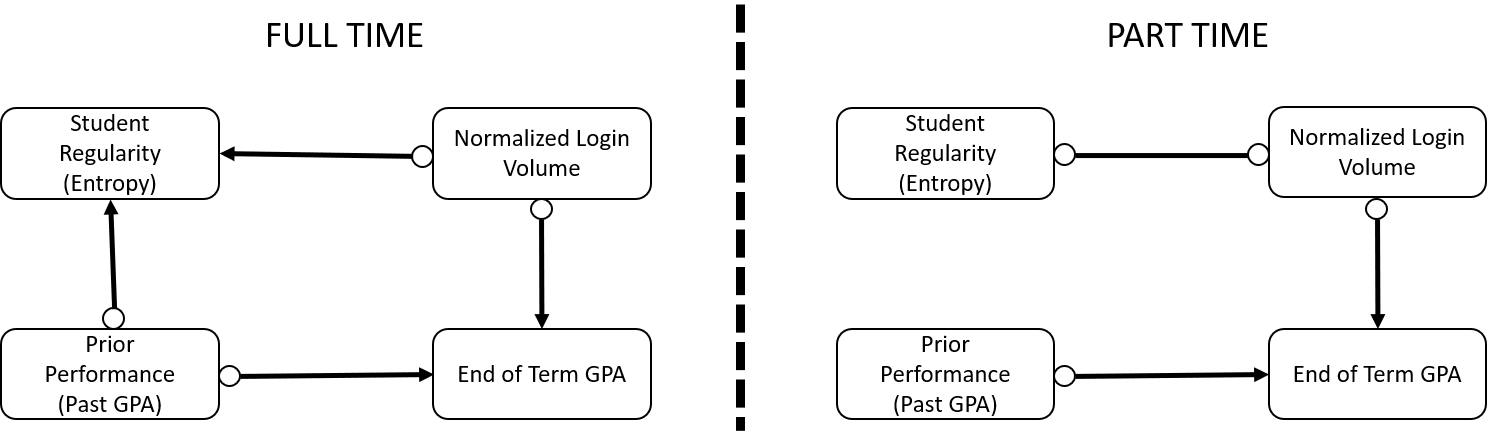}}
    \subfigure[]{\includegraphics[width=0.9\textwidth]{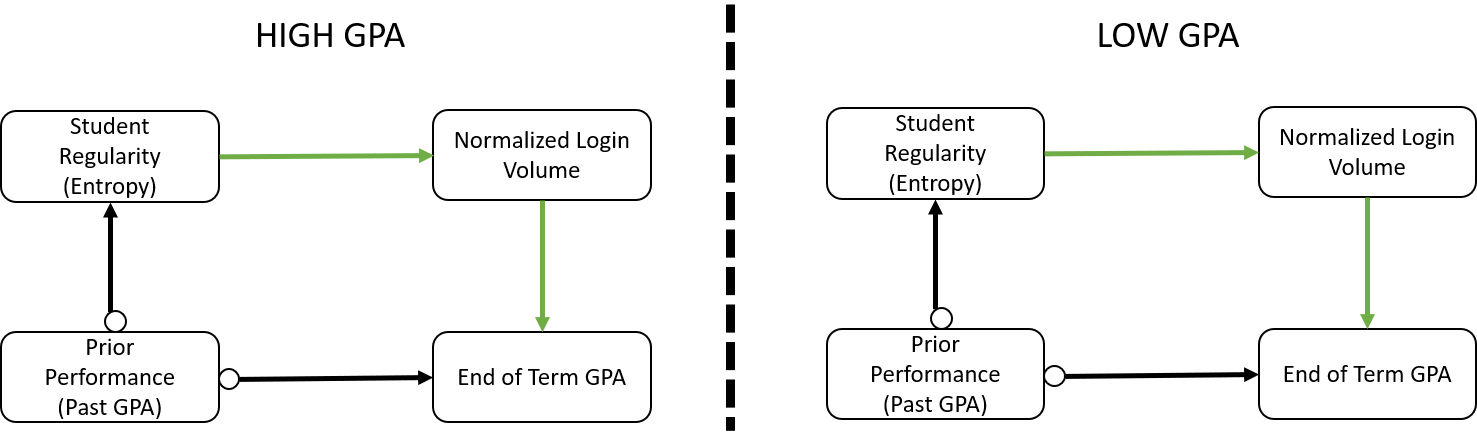}}
    \caption{(a) Causal relationships based on gender-specific dataset (b) Causal relationships based on student ethnicity (c) Causal relationships based on student enrollment type (Full-time/Part-time) (d) Causal relationships based on student GPA}
    \label{fig:fig9}
\end{figure}

In the next step, we divide the data into subsets based on gender, race, and GPA to study causal relationships based on these demographics. In line with our demographic based correlation studies in \ref{Result: Model Explanation}, We only focus on "login volume", "Student Regularity (Entropy)" and "Prior performance". The gender based feature weights related to mCCA are mentioned in Appendix tables \ref{tab:D.2} and \ref{tab:D.3}. 
From \ref{fig:fig9} (a), we can observe that there is no direct relationship between login variables and performance for different gender and race classifications. One difference between Male and Female gender is the impact of student prior performance on student login regularity.

There is a relationship between student regularity and their performance for Males, but for females, this does not exist. From \ref{fig:fig9} (b), we do not see any significant difference in relationships between students from Majority races and students from Minority races. The mCCA feature weights related to ethnicity based datasets are shown in Appendix tables \ref{tab:D.4} and \ref{tab:D.5}.
One interesting observation is the causal relationship between login volume and performance for minority race students, which we observed in our earlier work [210] with predictive analytics as well. The relationship between prior performance and regularity exists in full-time students but not part-time students, as shown in \ref{fig:fig9} (c). The corresponding mCCA weights for Full time and Part time students are shown in Appendix tables \ref{tab:D.6} and \ref{tab:D.7}.
There is a significant and direct relationship between login variables and student performance for students with varying bands of GPA ($< 2$ Low GPA \& $>3$ High GPA), as shown in \ref{fig:fig9} (d). The mCCA weights based on GPA are shown in Appendix Table \ref{tab:D.8} and Table \ref{tab:D.9}. 
Even though there are no clear relationships without confounders based on gender and race, it is still valuable to learn that a relationship exists between student logins and their performance.

\section{Discussion}
\label{Discussion}
With the increasing adoption of LMS systems in colleges and universities, there is growing interest in understanding student learning behaviors based on their interaction with these systems. Studies in Learning Analytics and Educational Data Mining have been focusing on developing models that predict student performance based on their interaction data captured by these systems. While those models provides certain insights on the correlational relationship between \textit{what students do}(as characterized by their LMS interactions) and \textit{what will happen in the future}(e.g. their academic performance), they may not be directly translated into actionable insights due to the inherent limitation of correlation analysis. Taking a \textbf{student-centric} view in analyzing LMS activities, we explored an analytical framework that can triangulate multiple aspect of students login behaviors including login volume, regularity and chronotype and quantify their relationships with academic performance. In order to elucidate the contributing  causal factors to students' learning outcome, we goes beyond the correlation analysis by employing a framework of causal discovery and inference. Here, we summarize a few key findings:

\subsection{Insights from Causal Analysis}
\label{Disc: Causal Analysis}

From the analysis in this study, we noted that student login volume is not only correlated, but also causally linked to students' academic performance. Moreover, it also revealed that login volume is influenced by student other self-regulation indicators such as logins regularity. This is an important finding as this shows that targeting an intervention that can increases student logins or their login regularity might improve their overall academic performance. The group-based study also demonstrated interesting and important heterogeneity of the impact of logins on students from different backgrounds. For example, we note the relationships between self-regulation variables and performance are stronger in students with lower GPA and also for students from minority groups. However, we found no significant gender difference in terms of the impact of LMS features on outcomes, nor noteable difference between regular and transfer students. These findings confirm that studies at the student level reveal valuable insights for various groups of students and can act as strong inputs for effective intervention development.



\subsection{Insights from Model Explanation on rich set of LMS behaviors }
\label{Disc: Model Explanation and Interpretation}
In this study, we focused on understanding the impact of a rich set of LMS behavior indicators on student performances. In addition to login volume, which was the primary focus of earlier studies, we also studied the impact of login regularity measured by entropy statistics on student performance by implementing LIME explanation, correlation, and linear regression methods. From our model explanation studies, we observed that students who login regularly into the LMS system have a positive relationship with performance improvement. The model explanation outcomes on this data also showed a positive relationship between increase in student login volumes and GPA. Even though this observation is accurate across all students, it has a slightly higher importance in students from minority races and student with GPA lower than 2 based on the data analyzed. One major limitation with LIME and other correlation-based methods is their lack of explanation of cause-and-effect relationship that is crucial for interventions. To overcome this, we are performed a causal analysis and discovery to study the relationship between LMS features and student performance.

\subsection{Insights from Chronotype Analysis}
\label{Disc: Chronotype Analysis}
One of the primary contributions of this study is the identification of student chronotypes based on their interaction variables. Earlier studies discussed the chronotype patterns in individuals and their impact on productivity \cite{diaz2008morningness,preckel2011chronotype,preckel2009ref}. These studies showed that individuals at younger ages are highly active in the morning, but this shifts to evening as they reach adolescence. The clustering methodology employed in this study showed a similar pattern where students in an undergraduate university showed high activity during the afternoon to evening hours. Prior studies also showed that these chronotypes have statistically significant relationships with academic achievement. However, these studies are primarily performed on a single domain or course \cite{zavgorodniaia2021morning}. Our analysis explicitly targets students as a single entity and aggregates their activity across all enrolled courses.The statistical findings in our work showed significant relationships between student demographics and chronotypes but not between chronotypes and their performance. In addition to this, we also performed a predictive modeling methodology to study if these chronotypes contribute to student performance prediction. The findings from this analysis revealed that student performance prediction stayed similar with and without the chronotype variables. These observations align with some earlier studies \cite{preckel2011chronotype,zavgorodniaia2021morning} that contradicted the finds of relationships between chronotypes and performance. Overall, research on the relationship between student performance and chronotypes needs more investigation as the findings between different studies are inconsistent.

\section{Limitation}
\label{Disc: Limitation}
There are also some significant limitations in this study. The data captured by LMS is only a snapshot of activity in students’ day-to-day lives. Student performance factors can be influenced by many other external factors like study environment, family background, and student perception towards a course. The login-based chronotypes discussed in this study are only a part of time management. Students' time management can be observed from multiple other factors like assignment submissions, time spent on course contents, and exploring time lag between lecture delivery and student reading content access. Another limitation is related to the dataset. This aggregate-level dataset only captures the student login information but not content-level information. Content level information is much more fine-grained and provides much more insights into what a student might be working on when they are logging into the system. One significant challenge with content-level information is their diversity based on course and instructor style. This will make it hard to extract student-level aggregated features.

\section{Conclusion}
\label{Conclusion}

To conclude, the causal analysis in this study strengthened our earlier findings that showed significant relationships between student login variables and their performance. The findings of this study are valuable to Educational Data Mining and Learning Analytics community as they support the design and development of interventions techniques based on LMS variables to improve student performance. As the data in this study is collected till the middle of the semester, the development and deployment of interventions at this stage provides valuable time to students for improvement and contribute to their academic achievement. In addition, This study explored a multi-prong methodology frameworks that involves model explanation and causal analysis, working together, they provide convincing evidence for interventions that targets students' LMS activity. The group-specific analysis further provide guidance on what specific student group to target. We envision that this type of methodological frameworks may stimulate many follow up work in this area to turn data into useful insights and further into action and intervention that has the promise to improve students' outcome. 

\newpage
\section{Appendix}

\begin{table}[h!]
    \caption{Composite variables with associated original features and weights estimated by mCCA for all students}\vspace*{1ex}
    \centering
    \begin{tabular}{l c c}
    \hline
     Composite Variable & Feature Name & Weights \\
     \hline
     \multirow{2}{4cm}{Normalized Login Volume} & Max Volume & 0.36\\
     & Median Volume & 0.93 \\
     \setlength{\parskip}{1ex}\\
     \multirow{2}{4cm}{Login Regularity (Entropy)} & Mean Entropy & -0.36\\
     & Median Volume & -0.93 \\
     \setlength{\parskip}{1ex}\\
     \multirow{2}{4cm}{Hourly Login Volumes} & H1 to H9 & \textless 0.18\\
     & H10 to H24 & 0.24 \\
     \setlength{\parskip}{1ex}\\
     \multirow{3}{4cm}{Hourly Time Spent} & H1 to H8 & 0.14 to 0.17\\
     & H9 to H21 & 0.21 to 0.26 \\
     & H22 to H24 & 0.19\\
     \setlength{\parskip}{1ex}\\
     \multirow{2}{4cm}{Weekday/Weekend Login Volumes} & WeekDay & 0.99\\
     & WeekEnd & 0.1 \\
     \hline
    \end{tabular}
    \label{tab:D.1}
\end{table}

\begin{table}[h!]
    \caption{Composite variables with associated original features and weights estimated by mCCA for MALE students}
    \centering
    \begin{tabular}{l c c}
    \hline
     Composite Variable & Feature Name & Weights\\
    \hline
    \multirow{5}{4cm}{Normalized Login Volume} & Min Volume & -0.3\\
    & Max Volume & -0.52\\
    & Mean Volume & -0.54\\
    & Median Volume & -0.55\\
    & Standard Deviation Volume & -0.2\\
    \setlength{\parskip}{1ex}\\
    \multirow{7}{4cm}{Login Regularity (Entropy)} & Min Entropy & 0.41\\
    & Max Entropy & 0.43\\
    & Mean Entropy & 0.57\\
    & Median Entropy & 0.57\\
    & Standard Deviation Entropy & 0.06\\
    & Skewness Entropy & -0.02\\
    & Kurtosis Entropy & 0.002\\
    \hline
    \end{tabular}
    \label{tab:D.2}
\end{table}

\begin{table}[ht!]
    \caption{Composite variables with associated original features and weights estimated by mCCA for FEMALE students}
    \centering
    \begin{tabular}{l c c}
    \hline
     Composite Variable & Feature Name & Weights\\
    \hline
    \multirow{2}{4cm}{Normalized Login Volume} & Mean & 0.11\\
    & Median & 0.99\\
    \setlength{\parskip}{1ex}\\
    \multirow{2}{4cm}{Login Regularity (Entropy)} & Mean & -0.99\\
    & Median & -0.11\\
    \hline
    \end{tabular}
    \label{tab:D.3}
\end{table}

\begin{table}[h!]
    \caption{Composite variables with associated original features and weights estimated by mCCA for Majority Race (White \& Asian) students}
    \centering
    \begin{tabular}{l c c}
    \hline
     Composite Variable & Feature Name & Weights\\
    \hline
    \multirow{2}{4cm}{Normalized Login Volume} & Mean & -0.11\\
    & Median & -0.99\\
    \setlength{\parskip}{1ex}\\
    \multirow{2}{4cm}{Login Regularity (Entropy)} & Mean & 0.99\\
    & Median & 0.11\\
    \hline
    \end{tabular}
    \label{tab:D.4}
\end{table}

\begin{table}[h!]
    \caption{Composite variables with associated original features and weights estimated by mCCA for Minority Race students}
    \centering
    \begin{tabular}{l c c}
    \hline
     Composite Variable & Feature Name & Weights\\
    \hline
    \multirow{2}{4cm}{Normalized Login Volume} & Mean & 0.11\\
    & Median & 0.99\\
    \setlength{\parskip}{1ex}\\
    \multirow{2}{4cm}{Login Regularity (Entropy)} & Mean & -0.11\\
    & Median & -0.99\\
    \hline
    \end{tabular}
    \label{tab:D.5}
\end{table}

\begin{table}[h!]
    \caption{Composite variables with associated original features and weights estimated by mCCA for Full-time students}
    \centering
    \begin{tabular}{l c c}
    \hline
     Composite Variable & Feature Name & Weights\\
    \hline
    \multirow{5}{4cm}{Normalized Login Volume} & Mean & -0.56\\
    & Median & -0.59\\
    & Min & -0.13 \\
    & Max & -0.53\\
    & Std & -0.07\\
    \setlength{\parskip}{1ex}\\
    \multirow{4}{4cm}{Login Regularity (Entropy)} & Mean & 0.62\\
    & Median & 0.62\\
    & Min & 0.30\\
    & Max & 0.37\\
    \hline
    \end{tabular}
    \label{tab:D.6}
\end{table}

\begin{table}[ht!]
    \caption{Composite variables with associated original features and weights estimated by mCCA for Part-time students}
    \centering
    \begin{tabular}{l c c}
    \hline
     Composite Variable & Feature Name & Weights\\
    \hline
    \multirow{2}{4cm}{Normalized Login Volume} & Max & -0.99\\
    & Median & -0.11\\
    \setlength{\parskip}{1ex}\\
    \multirow{2}{4cm}{Login Regularity (Entropy)} & Mean & 0.99\\
    & Median & 0.11\\
    \hline
    \end{tabular}
    \label{tab:D.7}
\end{table}

\begin{table}[h!]
    \caption{Composite variables with associated original features and weights estimated by mCCA for Low GPA ($\ll$2) students}
    \centering
    \begin{tabular}{l c c}
    \hline
     Composite Variable & Feature Name & Weights\\
    \hline
    \multirow{2}{4cm}{Normalized Login Volume} & Mean & 0.11\\
    & Median & 0.99\\
    \setlength{\parskip}{1ex}\\
    \multirow{2}{4cm}{Login Regularity (Entropy)} & Mean & -0.11\\
    & Median & -0.99\\
    \hline
    \end{tabular}
    \label{tab:D.8}
\end{table}

\begin{table}[h!]
    \caption{ Composite variables with associated original features and weights estimated by mCCA for High GPA ( $\geq3$ ) students}
    \centering
    \begin{tabular}{l c c}
    \hline
     Composite Variable & Feature Name & Weights\\
    \hline
    \multirow{2}{4cm}{Normalized Login Volume} & Mean & -0.36\\
    & Median & -0.93\\
    \setlength{\parskip}{1ex}\\
    \multirow{2}{4cm}{Login Regularity (Entropy)} & Mean & 0.93\\
    & Median & 0.36\\
    \hline
    \end{tabular}
    \label{tab:D.9}
\end{table}


\clearpage
\bibliography{LA-bibfile} 
\bibliographystyle{acmtrans}

\end{document}